\documentclass[cernpreprint, PAPER, atlasdraft=false, texlive=2023, UKenglish, texmf, orcidlogo]{atlasdoc}

\usepackage{atlaspackage}
\usepackage{atlasbiblatex}

\usepackage{atlasphysics}
\graphicspath{{logos/}{figures/}}

\usepackage{ANA-HDBS-2019-19-PAPER-defs}

\AtlasTitle{Search for short- and long-lived axion-like particles in $H\rightarrow a a \rightarrow 4\gamma$ decays with the ATLAS experiment at the LHC}

\AtlasAbstract{%
Presented is the search for anomalous Higgs boson decays into two axion-like particles (ALPs) using the full Run~2 data set of $140\,\text{fb}^{-1}$ of proton-proton collisions at a centre-of-mass energy of $13\,\text{TeV}$ recorded by the ATLAS experiment. The ALPs are assumed to decay into two photons, providing sensitivity to recently proposed models that could explain the $(g-2)_\mu$ discrepancy. This analysis covers an ALP mass range from $100\,\text{MeV}$ to $62\,\text{GeV}$ and ALP-photon couplings in the range $10^{-7}\,\text{TeV}^{-1}<C_{a\gamma\gamma}/\Lambda<1 \,\text{TeV}^{-1}$, and therefore includes signatures with significantly displaced vertices and highly collinear photons. No significant excess of events above the Standard Model background is observed. Upper limits at 95\% confidence level are placed on the branching ratio of the Higgs boson to two ALPs in the four-photon final state, and are in the range of $ 10^{-5}$ to $3\times 10^{-2}$, depending on the hypothesized ALP mass and ALP-photon coupling strength.
}

\AtlasRefCode{HDBS-2019-19}

\PreprintIdNumber{CERN-EP-2023-202}

\AtlasJournal{\EPJC}
\AtlasJournalRef{\EPJC 84 (2024) 742}
\AtlasDOI{10.1140/epjc/s10052-024-12979-0}

\hypersetup{pdftitle={ATLAS document},pdfauthor={The ATLAS Collaboration}}

\begin{document}

\maketitle

\tableofcontents

%
%

%

%
\section{Introduction}
\label{sec:intro}

Axion-like particles (ALPs), or more generally light (pseudo) scalars, are represented as gauge-singlets beyond the Standard Model (SM) that can couple to the 125~\GeV\ Higgs boson~\cite{HIGG-2012-27,CMS-HIG-12-028}, and appear in many well-motivated extensions of the SM, e.g., the next-to-minimal supersymmetric standard model~\cite{Dobrescu:2000yn,Ellwanger:2003jt}. These include models that address the baryon asymmetry of the universe~\cite{Profumo:2007wc,Blinov:2015sna}, offer a solution to the naturalness problem~\cite{Craig:2015pha,Curtin:2015fna}, or provide insights into the nature of dark matter~\cite{Silveira:1985rk,Pospelov:2007mp,Draper:2010ew,Ipek:2014gua,Martin:2014sxa,Boehm:2014hva}. ALPs and light bosons produced in Higgs boson decays could also be mediators to dark sectors that do not otherwise couple to the SM~\cite{Curtin:2013fra}.

A combination of ATLAS measurements of Higgs boson properties using $139\,\ifb$ of data constrains the branching ratios into invisible and undetected states to be $\mathcal{B}(H\rightarrow\text{invisible})<10.7\%$ and $\mathcal{B}(H\rightarrow\text{undetected})<12\%$ at 95\% confidence level (CL)~\cite{ATLAS:2023tkt, ATLAS:2022vkf}.
Combined measurements of Higgs boson couplings performed by the CMS Collaboration using $138\,\ifb$ of data set upper limits of $\mathcal{B}(H\rightarrow\text{invisible})<16\%$ and $\mathcal{B}(H\rightarrow\text{undetected})<16\%$ at 95\% CL~\cite{CMS:2022dwd}.
These results allow potentially large branching fractions into beyond-the-standard-model (BSM) particles, $\mathcal{B}(H\rightarrow\text{BSM})$, such as ALPs.

In  Ref. \cite{Bauer:2017ris} it has been argued that if the ALP, denoted as $a$ in the following, couples to at least some SM particles with couplings of order (0.01-1)TeV$^{-1}$, its mass must be above 1 MeV.
Taking into account the possibility of a long-lived ALP, large regions of so far unconstrained parameter space can be explored by searches for exotic, on-shell Higgs boson decays into two ALPs.
In particular, this includes the parameter space in which ALPs can explain the observed discrepancy between the measurement \cite{Muong-2:2021ojo, Muong-2:2006rrc, Muong-2:2023cdq} and the theoretical prediction \cite{Aoyama:2020ynm, RBC:2018dos} of  the anomalous magnetic moment of the muon \cite{Bauer:2017ris}.
It was suggested that subsequent ALP decays into photons provide unprecedented sensitivity to the ALP-photon couplings in the mass region above a few MeV, even if the relevant ALP-photon couplings are loop suppressed and the $a \to \gamma\gamma$ branching ratios are significantly less than 100\% \cite{Bauer:2017ris}.

This paper presents a search for decays of the 125~\GeV\ Higgs boson into two ALPs in proton--proton ($pp$) collisions at the LHC~\cite{Evans:2008zzb}.
The search is sensitive to events where each $a$-boson decays into two photons.
For the first time, a dedicated search for long-lived $a\to\gamma\gamma$ decays with a significantly displaced vertex within the tracking system of the ATLAS detector is performed, allowing a large region of the parameter space of the ALP-photon coupling $C_{a\gamma\gamma}$ to be probed.
Previous searches for $H\to aa \to 4\gamma$ signatures were performed by the ATLAS~\cite{ATLAS:2015rsn} and CMS Collaborations~\cite{CMS:2022fyt, CMS:2022xxa}, but assumed promptly decaying ALPs and hence were valid only up to decay lengths of a few centimeters.  The decay length scales with
$\tau_a \propto \Lambda ^{2} / \left( m_a^3 |C_{a\gamma\gamma}|^{2} \right)$, where $m_a$ is the ALP mass and $\Lambda$,
the new physics scale, is assumed to be in the TeV range \cite{Bauer:2017ris}.
The limits obtained here for ALP masses $m_a>15\,\GeV$ are about one order of magnitude more stringent than previous ATLAS analyses using $8\,\TeV$ data,
and reach sensitivies similar to or slightly better than
previous analyses from CMS using $132\,\ifb$ of $\sqrt{s} = 13\,\TeV$ data.

The paper is structured as follows. A brief discussion of the ATLAS detector and an overview of the Monte Carlo (MC) samples and data sets used are given in Sections~\ref{sec:detector} and \ref{sec:datasets}.
Object reconstruction and the event selection are described in Section~\ref{sec:eventobjects}, where special focus is given to the reconstruction of collimated photon signatures
and to the categorization of the final state topologies that can be reconstructed in the detector.
The background estimates are discussed in Section~\ref{sec:background}, followed by a description of the dominant systematic uncertainties in Section~\ref{sec:systematics}, in particular those that involve the reconstruction of photons with a displaced production vertex. The statistical interpretation and the final results are summarized in Section~\ref{sec:results}. The paper closes with a brief conclusion in Section~\ref{sec:conclusion}.


\section{The ATLAS Detector}
\label{sec:detector}
The ATLAS detector~\cite{PERF-2007-01} at the LHC covers nearly the entire solid angle around the collision point.\footnote{\AtlasCoordFootnote}
It consists of an inner tracking detector surrounded by a thin superconducting solenoid, electromagnetic and hadron calorimeters,
and a muon spectrometer incorporating three large superconducting air-core toroidal magnets.

The inner-detector system (ID) is immersed in a \qty{2}{\tesla} axial magnetic field
and provides charged-particle tracking in the range of \(|\eta| < 2.5\).
The high-granularity silicon pixel detector covers the vertex region and typically provides four measurements (hits) per track,
the first hit normally being in the insertable B-layer (IBL) installed before Run~2~\cite{ATLAS-TDR-19,PIX-2018-001}.
It is followed by the silicon microstrip tracker (SCT), which usually provides eight measurements per track.
These silicon detectors are complemented by the transition radiation tracker (TRT),
which enables radially extended track reconstruction up to \(|\eta| = 2.0\).
The TRT also provides electron identification information
based on the fraction of hits (typically 30 in total) above a higher energy-deposit threshold corresponding to transition radiation.

The calorimeter system covers the pseudorapidity range of \(|\eta| < 4.9\).
Within the region \(|\eta|< 3.2\), electromagnetic calorimetry is provided by barrel and
endcap high-granularity lead/liquid-argon (LAr) calorimeters,
with an additional thin LAr presampler covering \(|\eta| < 1.8\)
to correct for energy loss in material upstream of the calorimeters.
Hadron calorimetry is provided by the steel/scintillator-tile calorimeter,
segmented into three barrel structures within \(|\eta| < 1.7\), and two endcap copper/LAr hadron calorimeters up to \(|\eta| < 3.2\).
The solid angle coverage is completed with forward copper/LAr and tungsten/LAr calorimeter modules
optimised for electromagnetic and hadronic energy measurements, respectively.

The muon spectrometer (MS) comprises separate trigger and
high-precision tracking chambers that measure the deflection of muons in the magnetic field generated by the superconducting air-core toroidal magnets.
The field integral of the toroids ranges between \num{2.0} and \qty{6.0}{\tesla\-\metre}
across most of the detector.
Three layers of precision chambers, each consisting of layers of monitored drift tubes, cover the region \(|\eta| < 2.7\),
complemented by cathode-strip chambers in the forward region.
The muon trigger system covers the range of \(|\eta| < 2.4\) with resistive-plate chambers in the barrel, and thin-gap chambers in the endcap regions.

Events of interest for this analysis are selected using dedicated di-photon triggers. The first-level of the trigger system is implemented in custom hardware,
followed by a high-level trigger~\cite{TRIG-2016-01} where further selections are made by algorithms implemented in software.
The first-level trigger accepts events from the \SI{40}{\MHz} bunch crossings at a rate below \SI{100}{\kHz},
which the high-level trigger further reduces in order to record events to disk at about \SI{1}{\kHz}.

An extensive software suite~\cite{ATL-SOFT-PUB-2021-001} is used in MC simulation, in the reconstruction
and analysis of real and simulated data, in detector operations, and in the trigger and data acquisition
systems of the experiment.


\section{Data Set and Simulated Events\label{sec:datasets}}

This analysis uses $pp$ collision data collected by the ATLAS experiment from 2015 to 2018 with a centre-of-mass energy of $\sqrt s =13$~\TeV. After data quality requirements~\cite{DAPR-2018-01} the full data set corresponds to an integrated luminosity of $140.1 \pm 1.2$~fb$^{-1}$~\cite{ATLAS:2022hro}. The lowest-threshold unprescaled di-photon triggers are used to select events for further analysis, as discussed in Section~\ref{sec:events}.

The main background contributions stem from multijet processes, where jets might be misidentified as photons, as well as Standard Model multiphoton processes \cite{ATLAS:2017lpx}.
However, MC simulation do not capture the full background shape observed in data and therefore a fully data-driven background estimation has been employed.
Simulated events are used to validate the data-driven background estimation techniques described in Sections~\ref{sec:background:2y} -- \ref{sec:background:3y}, and to model Higgs boson decays into photons for both signal and background processes.
Background events from continuum $3\gamma$, $4\gamma$ and $\gamma\gamma$ production, neglecting any interference effects with the $H\rightarrow\gamma\gamma$
decay, are generated using Sherpa 2.2.8 and Sherpa 2.2.4~\cite{Gleisberg:2008ta,Bothmann:2019yzt,Hoeche:2009rj}, respectively,  with the AZNLO set of tuned parameters~\cite{STDM-2012-23} and the
NNPDF30~\cite{Ball:2014uwa} set of parton distribution functions (PDF).
Production of the Higgs boson through  gluon--gluon fusion (ggF) and vector-boson fusion (VBF) processes is modelled at next-to-leading order (NLO) using \textsc{Powheg-Box} v2~\cite{Nason:2004rx,Frixione:2007vw,Alioli:2010xd} interfaced with \textsc{Pythia}
8.186~\cite{Sjostrand:2014zea} using the AZNLO tune and NNPDF30 PDF.
The change in selection efficiency between ggF and VBF and other Higgs boson production mechanisms is at the percent level, so its impact is negligible. Thus, all signal ($H\rightarrow aa \rightarrow \gamma\gamma\gamma\gamma$) samples use the ggF Higgs boson production mechanism and have their cross-sections scaled to the total Higgs boson production cross-section predicted by a
next-to-next-to-next-to-leading-order QCD calculation with NLO electroweak corrections applied~\cite{deFlorian:2016spz,Anastasiou:2016cez,Actis:2008ug,Anastasiou:2008tj}.

Signal samples are produced for ALP masses in the range of $0.1\,\GeV$ -- $62\,\GeV$, with a spacing of $0.1\,\GeV$ in the range of $0.1\,\GeV$ -- $0.5\,\GeV$, a spacing of $0.5\,\GeV$ in the range of $0.5\,\GeV$ -- $5\,\GeV$ and a spacing of $1\,\GeV$ in the range of $5\,\GeV$ -- $62\,\GeV$, assuming $\Lambda$ = 1~\TeV.
The full set of mass points is simulated for four different values of the coupling $C_{a\gamma\gamma}$: $1$, $0.01$, $5\times10^{-4}$ and $10^{-5}$. Samples of simulated signal events are generated using \textsc{Powheg-Box} v2 at NLO for the Higgs production and \textsc{Pythia} 8.212 to simulate the decay into ALPs and the subsequent decay into photons employing a generic 2HDM model \cite{Grzadkowski:2018ohf}. For hadronisation and parton showering, \textsc{Pythia} 8.212 is used as well, with the AZNLO tune and the CTEQ6L1 PDF set~\cite{Stump:2003yu}.

Different pile-up conditions due to additional $pp$ interactions in the same and neighbouring bunch crossings are simulated by overlaying the hard-scattering event with inelastic $pp$ events generated by \PYTHIA[8.186] using the \NNPDF[2.3lo] PDF set~\cite{Ball:2012cx} and the A3 tune~\cite{ATL-PHYS-PUB-2016-017}.
Differences between the simulated and observed distributions of the number of interactions per bunch crossing are corrected by reweighting the simulated events to match the data distribution.

All MC samples use a full simulation of the ATLAS detector~\cite{SOFT-2010-01} based on \GEANT~\cite{Agostinelli:2002hh} to reproduce the detector response. Corrections are applied to the simulated events to match the photon selection efficiencies, energy scale and resolution to those determined in data.


\section{Object Reconstruction and Event Selection\label{sec:eventobjects}}

The experimental signature of $H\rightarrow aa \rightarrow 4\gamma$ depends significantly on the ALP mass and on its coupling to photons, $C_{a\gamma\gamma}$.
ALP masses below $3.5\,\GeV$ dominantly yield collimated photon signatures, which are reconstructed as one energy cluster in the electromagnetic calorimeter.
The photons originating from the decays of axion-like particles with higher masses can typically be resolved by the ATLAS calorimeter system and the ATLAS identification algorithms.
The $C_{a\gamma\gamma}$ coupling determines the lifetime and hence the  distance the ALPs travel after being produced, so the search for $H\rightarrow aa \rightarrow 4\gamma$ is divided into two sub-searches: a search for promptly decaying ALPs corresponding to couplings $C_{a\gamma\gamma} \ge 0.1$  and a search for long-lived ALPs that is optimised for smaller couplings. Coupling values smaller than $C_{a\gamma\gamma}=10^{-7}$ cannot be probed within this analysis since they yield ALP lifetimes that imply a decay outside of the detector volume for all ALP masses considered.

\subsection{Object Reconstruction}

Photons are reconstructed from topologically connected clusters~\cite{ATL-PHYS-PUB-2017-022} of energy deposits in the electromagnetic calorimeter in the region $|\eta| < 2.5$ where a distinction to electrons is provided by the tracking system.
Only photon candidates within  $|\eta| < 2.37$ are used in this analysis to allow the usage of a track based isolation variable.
The transition region between the barrel and endcap electromagnetic calorimeters, $1.37 < |\eta| < 1.52$, is excluded due to poor resolution of the calorimeter in this area.
The analysis does not distinguish between photons converting into $e^+e^-$ pairs and unconverted photons.

Photon candidates are defined by the lateral shower profile of the energy deposits in the first and second electromagnetic calorimeter layers and by the fractional energy leakage into the hadron calorimeter.
The analysis uses different photon identification criteria for different regions of the signal parameter space.
Standard `Tight' and `Loose' photon identification criteria~\cite{ATLAS:2023dxj, EGAM-2018-01},
which are tuned for converted and unconverted photons separately, are applied in final states where individual photons can be separately reconstructed. Axion-like particles with small masses, e.g., $m_a <3.5\,\GeV$, decay predominantly into a pair of highly collimated photons, which are reconstructed as single photon object. For these merged photon objects, the standard photon identification is not efficient.
Hence, a neural network (NN)-based classification approach is developed. A first classifier is trained to separate real photon signatures, single or collimated, from ‘fake photons’ that come from multijet background.
A second classifier is then trained to separate single-photon signatures from collimated signatures. Both classifiers use eight shower-shape variables relevant for photon identification~\cite{EGAM-2018-01}.
The training data sets are based on single- and collimated-photon signatures from simulation and fake-photon candidates from data; the fake-photon candidates are selected by inverting the requirement on several isolation observables.
If a photon passes a minimum threshold of the output neurons of both classifiers, it is labelled as a `merged' photon. The performance of the classifier is shown in Fig. \ref{fig:NNPhotonID}.
Due to coarse pointing resolution of of the calorimeter it is not possible to reconstruct displaced secondary vertices from ALP decays.

\begin{figure}[thb]
\centering
\subfloat[]{
\includegraphics[width=0.48\textwidth]{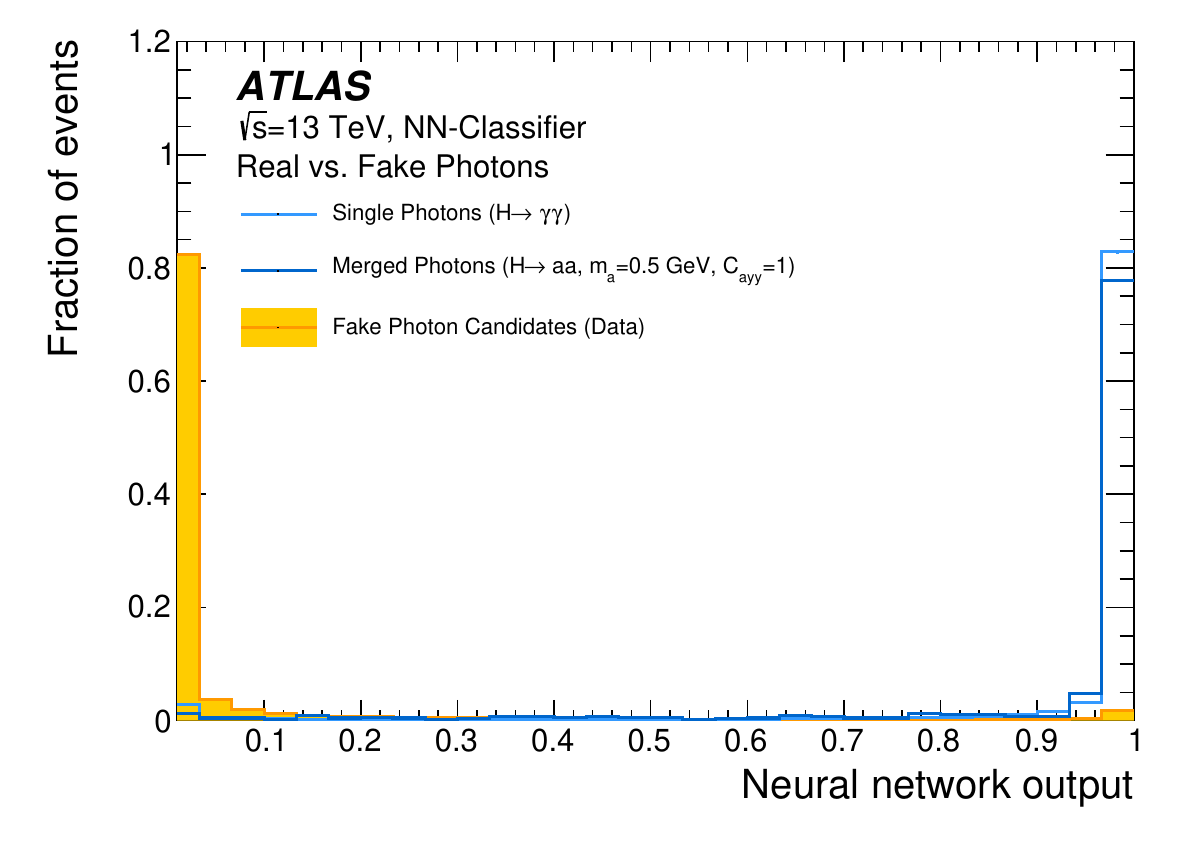}
} \hfill
\subfloat[]{
\includegraphics[width=0.48\textwidth]{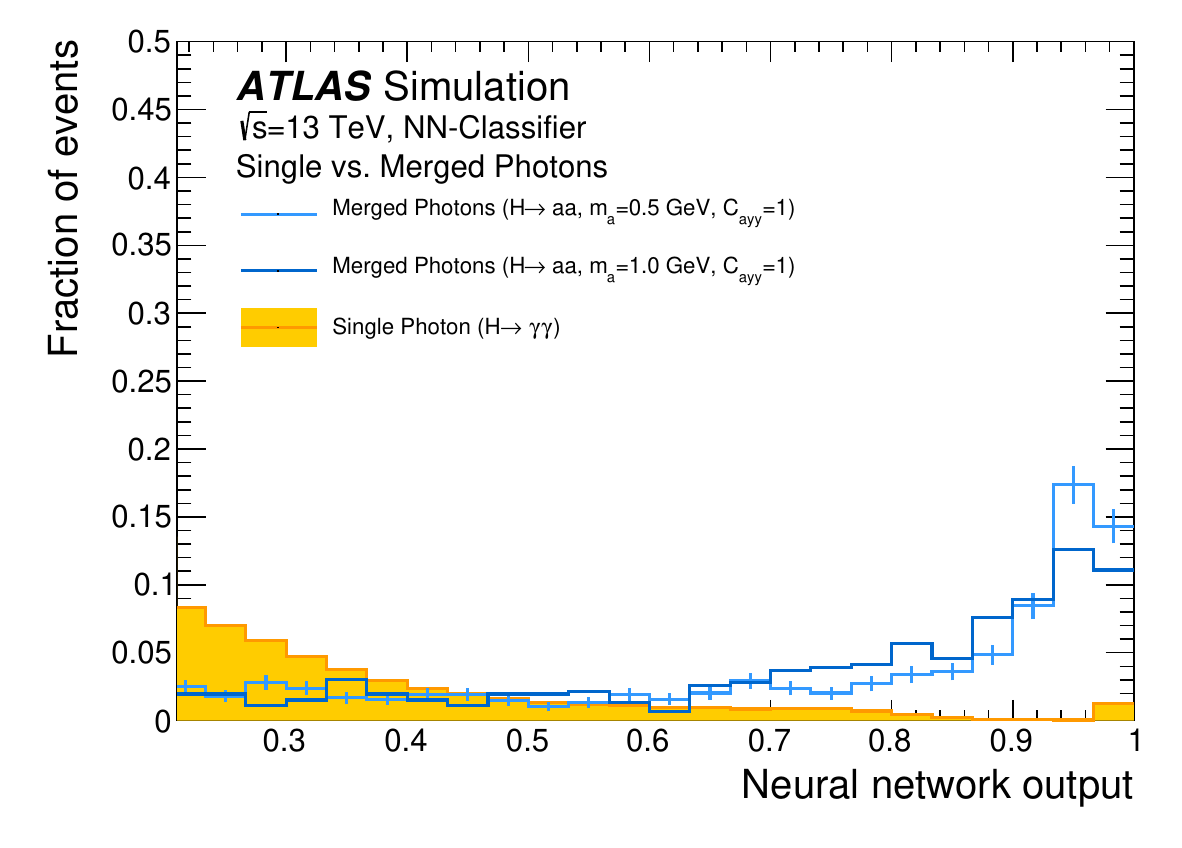}
} \hfill
\caption{Output of the neural network classifier to distinguish (a) real from fake photons, and (b) single from merged photons.}
\label{fig:NNPhotonID}
\end{figure}

The photon energy is reconstructed using the nominal ATLAS reconstruction and calibration procedures~\cite{ATLAS:2023mnw}.
The energy of photon candidates that are identified as collimated photon signatures is corrected by adding the measured cluster energies in the calorimeter within a cone of $\Delta R = 0.2$ around the photon, if not yet associated with the photon.

To further improve the rejection of misidentified photons, a track-based  variable $p_{\textrm{T}}^{\textrm{iso}}$ is defined as the scalar sum of the transverse momenta of all tracks with transverse momentum ($\pT$) above 1~\GeV{} that originate from the primary vertex and are within a cone of $\Delta R = 0.2$ around the photon candidate with transverse energy $E^{\gamma}_{\textrm{T}}$.
Isolated photons must have $p_{\textrm{T}}^{\textrm{iso}}/E^{\gamma}_{\textrm{T}}<0.05$.
A calorimeter-based isolation requirement is not used due to its large signal rejection for collimated photon signatures.

\subsection{Event Selection and Categorization \label{sec:events}}

Events are selected using one of two di-photon triggers~\cite{TRIG-2018-05}, each of which requires two reconstructed photon candidates. The first trigger requires a minimum transverse energy of 35~\GeV\ for the leading photon and 25~\GeV\ for the subleading photon where both photons must satisfy the online Loose or Medium identification criteria ~\cite{TRIG-2018-05}, depending on the data-taking period. The alternative trigger requires
that each photon satisfy the Tight identification criteria and have a transverse energy of at least 22~\GeV.
The terms `leading' and `subleading' refer to the photon candidate with the highest and second-highest transverse energy, respectively.
Photon candidates in the event must have a minimum transverse energy corresponding to the trigger threshold for the leading and subleading photons and at least 15~\GeV\ for any additional photon.
For $H\to aa\to 4\gamma$ events with  $C_{a\gamma\gamma}=1$ and at least two selected photon candidates, the average trigger efficiency is larger than 60\% for all ALP masses.
This still holds for $C_{a\gamma\gamma} = 10^{-5}$
and $m_a > 50\,\GeV$, while for smaller masses the trigger efficiency decreases down to $30\,\%$. The trigger efficiency drops for decreasing ALP masses as collimated photon signatures often fail the isolation requirement at trigger level, which is based on calorimeter information rather than tracks.

Signal events are fully reconstructed if all four photons are detected.
However, some of the photons might be out of detector acceptance and some might not be reconstructed.
The lifetime of the ALPs depends on their mass and coupling to photons, as shown in Sec. \ref{sec:intro}. Some examples are listed here: assuming a coupling of $C_{a\gamma\gamma}  = 10^{-4}$ a mass of $m_a = 1\,\GeV$ ($m_a = 10\,\GeV$) yields a lifetime of $c\tau = 25\,\text{m}$ ($c\tau = 25\,\text{mm}$). For $C_{a\gamma\gamma}  = 10^{-5}$ the lifetime increases by a factor of hundred.
Missing photons as well as pairs of collimated photons being reconstructed as one photon result in fewer reconstructed photons.
At least two reconstructed photon candidates are required for the analysis.
Further classification is based on the number and types of reconstructed photons.

Each event is classified into one of the five categories according to their experimental signature, in the following order:
1) events with four reconstructed photons, where all photons satisfy the Loose identification and at least one satisfies the Tight identification, fall into the four-single (4S) category.
2) events with three reconstructed photons, where all photons satisfy the Tight identification, fall into the three-single (3S) category.
3) events with two merged photon candidates fall into the two-merged (2M) category. Additional loose photon candidates are likely to originate from background processes and are ignored, with negligible impact on the rate of falsely identified signal events.
4) events with two photon candidates, where one satisfies the merged classification and
the other satisfies the Loose identification, fall into the one-merged-one-single (1M1S) category.
This category accepts events where the merged classification of two photons is not efficient.
5) events with exactly two photons that satisfy the Tight identification but without any further photon candidates that satisfy the Loose identification fall into the two-single category (2S), which is dominated by events from the $H\rightarrow\gamma\gamma$  process.

Only the most sensitive categories are used for the ALP search.
For long-lived ALPs with masses $m_a\ge 3.5\,\GeV$ the 3S and 4S categories
provide the most sensitivity.
Long-lived ALPs with $m_a<3.5\,\GeV$ yield collimated photon signatures where the 2M, and 1M1S categories
provide the largest sensitivity.

The 3S and 4S categories allow for the reconstruction of the ALP mass, $m_a$, since at least one photon pair stems from the decay of an ALP for the signal process.
Separate neural networks were trained for the three- and four-photon categories to select the correct photon pairing(s).
The inputs to the networks are the invariant masses of all photon pair combinations and differences in their transverse energies and directions.
The training sample consists of both correct and wrong photon combinations.
The combinations are based on MC signal samples for all ALP masses and couplings. In the 3S category, the invariant mass of the photon pair that is predicted to stem from the same mother particle is defined as the reconstructed ALP mass $m^{\text{reco}}_a$, while the average of both resulting invariant masses is defined as $m^{\text{reco}}_a$ in the 4S category.

The signal region selection uses the invariant mass of all photon candidates, denoted by $m^{\text{reco}}_{\text{inv}}$, which is expected to peak around the Higgs boson mass for signal processes.
The signal regions for all categories and searches are defined to contain at least 90\% of the reconstructed signal contribution in simulated data.
For the two-photon categories (2M, 1M1S and 2S), the signal region is defined by the invariant mass requirement $115\,\GeV < m^{\text{reco}}_{\text{inv}} < 130\,\GeV$ around the Higgs boson mass.
The 3S and 4S categories also require the reconstructed ALP mass $m^{\text{reco}}_a$ to fall within a window around the generated ALP mass, whose size varies depending on the mass point being tested.
In general, the requirements on $m^{\text{reco}}_{\text{inv}}$ are looser in the 3S category since one photon typically escapes detection, leading to a wider $m^{\text{reco}}_{\text{inv}}$ spectrum.
The $m^{\text{reco}}_a$ distribution also enables the definition of a control region in data by inverting the mass requirement on $m^{\text{reco}}_a$.

In the search for promptly decaying ALPs only the 4S category is used when estimating the limits on the signal process, as it has by far the largest sensitivity.
In this case the category is defined by tightening the selection cuts so that at least three out of the four photons satisfy the Tight identification requirements and is labelled $4S_p$. Promptly decaying ALPs are only considered for $m_a > 5$~\GeV, where a significant increase in sensitivity over the search for long-lived ALPs is observed.
Table \ref{tab:SignalRegions} summarizes the signal region definitions for all ALP mass hypotheses m$_a$.

\begin{table}[ht]
\small
\begin{center}
\begin{tabular}{l|c|c}
\hline
Model Parameters                    &  \multicolumn{2}{c}{Signal Region Definition}  \\
\hline
\multicolumn{3}{c}{Long-lived ALP Search:  $C_{a\gamma\gamma}<0.1$ } \\
\hline\hline
&  \multicolumn{2}{c}{2M, 1M1S and 2S Categories}  \\
\hline
0.1~\GeV$\leq m_a<3.5$~\GeV                    & \multicolumn{2}{c}{115~\GeV $<m^{\text{reco}}_{\text{inv}}<$ 130~\GeV } \\
\hline
& 3S Category  & 4S Category  \\

\hline
3.5~\GeV$\leq m_a<10$~\GeV                    & 105~\GeV $<m^{\text{reco}}_{\text{inv}}<$ 130~\GeV  & 120~\GeV $<m^{\text{reco}}_{\text{inv}}<$ 130~\GeV \\
& 0~\GeV $<m^{\text{reco}}_a<$ 10~\GeV & 0~\GeV $<m^{\text{reco}}_{a}<$ 12~\GeV \\
\hline
10~\GeV$\leq m_a<25$~\GeV                    & 100~\GeV $<m^{\text{reco}}_{\text{inv}}<$ 125~\GeV  & 120~\GeV $<m^{\text{reco}}_{\text{inv}}<$ 130~\GeV \\
& 6~\GeV $<m^{\text{reco}}_{a}<$ 26~\GeV & 8~\GeV $<m^{\text{reco}}_{a}<$ 28~\GeV \\
\hline
25~\GeV$\leq m_a<40$~\GeV                    & 100~\GeV $<m^{\text{reco}}_{\text{inv}}<$ 125~\GeV  & 120~\GeV $<m^{\text{reco}}_{\text{inv}}<$ 130~\GeV \\
& 20~\GeV $<m^{\text{reco}}_{a}<$ 40~\GeV & 23~\GeV $<m^{\text{reco}}_{a}<$ 43~\GeV \\
\hline
40~\GeV$\leq m_a\le 62$~\GeV                    & 90~\GeV $<m^{\text{reco}}_{\text{inv}}<$ 115~\GeV  & 120~\GeV $<m^{\text{reco}}_{\text{inv}}<$ 130~\GeV \\
& 30~\GeV $<m^{\text{reco}}_{a}<$ 65~\GeV & 38~\GeV $<m^{\text{reco}}_{a}<$ 65~\GeV \\
\hline
\multicolumn{3}{c}{Prompt ALP Search:  $0.1<C_{a\gamma\gamma}<1$	 } \\
\hline\hline
&   		\multicolumn{2}{c}{$4S_p$ Category}  \\
\hline
5~\GeV$\leq m_a<25$~\GeV   		&   	\multicolumn{2}{c}{120~\GeV $<m^{\text{reco}}_{\text{inv}}<$ 130~\GeV} 	\\
& 	 	\multicolumn{2}{c}{$|m_a-m^{\text{reco}}_{a}| < 1$~\GeV 	}	\\

\hline
25~\GeV$\leq m_a<40$~\GeV   		&   	\multicolumn{2}{c}{120~\GeV $<m^{\text{reco}}_{\text{inv}}<$ 130~\GeV} 	\\
& 	 	\multicolumn{2}{c}{ $|m_a-m^{\text{reco}}_{a}| < 2$~\GeV 		}\\

\hline
40~\GeV$\leq m_a<50$~\GeV   		&  	\multicolumn{2}{c}{120~\GeV $<m^{\text{reco}}_{\text{inv}}<$ 130~\GeV 	}\\
&  	\multicolumn{2}{c}{ $|m_a-m^{\text{reco}}_{a}| < 3$~\GeV 		}\\

\hline
50~\GeV$\leq m_a<55$~\GeV   		& 			  	\multicolumn{2}{c}{ 120~\GeV $<m^{\text{reco}}_{\text{inv}}<$ 130~\GeV 	}\\
& 				 	\multicolumn{2}{c}{ $|m_a-m^{\text{reco}}_{a}| < 5$~\GeV 		}\\

\hline
55~\GeV$\leq m_a\le 62$~\GeV   		& 			  	\multicolumn{2}{c}{ 120~\GeV $<m^{\text{reco}}_{\text{inv}} < 130$~\GeV 	}\\
& 				 	\multicolumn{2}{c}{ $|m_a-m^{\text{reco}}_{a}| < 8$~\GeV 		}\\

\hline
\end{tabular}
\caption{Definition of the signal region for different event categories in the prompt and long-lived search\label{tab:SignalRegions}.}
\end{center}
\end{table}

%
%

%
%
%
%
%
%
%
%
%
%
%
%
%
%
%
%
%
%
%
%
%
%
%
%
%
%
%
%
%
%
%
%
%
%
%
%
%
%
%
%
%
%
%
%


\section{Background Estimation\label{sec:background}}

The main background contributions stem from multijet processes, where jets might be misidentified as photons, as well as Standard Model multiphoton processes.
However, as the simulation does not capture the full background shape observed in data we rely on a fully data driven background estimate.

The signal is expected to peak in $m^{\text{reco}}_{\text{inv}}$, the invariant mass of the selected photons, near the Higgs boson mass, $m_H=125\,\GeV$, in all categories and searches.
The $m^{\text{reco}}_{\text{inv}}$ sidebands are used for the background estimate in the signal region
and for the estimation of spurious signal effects.

\subsection{Two-Photon Final States in the Search for Long-Lived Axion-Like Particles\label{sec:background:2y}}

The distribution of $m^{\text{reco}}_{\text{inv}}$ in all two-photon categories (2M, 1M1S, 2S) is fitted over a mass range from 100 to 150~\GeV, region excluding the signal region. A suitable fitting function should describe the data in the sidebands, provide an unbiased estimate of the background in the signal region, and produce small uncertainties on the yields of spurious signals.
This is ensured by defining validation regions that use similar requirements to those of the signal region but reject signal events.
Simulated samples of two-photon continuum processes and a data-driven validation region are studied. The latter is defined using the nominal signal selection and classification, but inverting the isolation cut on the photon candidates, thus yielding a multijet-enhanced data sample.\footnote{It is ensured that there is no overlap with the training data of the NN-based photon identification classifiers.}
These validation regions are chosen to cover
any scenario between a background composed purely of di-photon events and a background consisting of mis-identified multijet events.

A Landau function gives unbiased background estimates in the signal region of both validation samples and provides $\chi^2$ per number of degrees of freedom around unity ($\chi^2/{\text{ndf}} \approx 1$) in all sideband regions for the 2S and 1M1S categories.
It is expected that the background shape in the 2M category differs from the 2S and 1M1S categories, due to the different background composition after the requirements on the NN-based classifiers. A second-order polynomial provides a good description of all validation regions of the 2M category. Variations of the background estimates using different fitting functions and different fitting ranges are used to define systematic uncertainties and are discussed in Section~\ref{sec:systematics}.

The $H\rightarrow\gamma\gamma$ process contributes not in the sideband but in the signal region as irreducible background, significantly in the 2S category and to a negligible extent in the 2M and 1M1S categories. Its contribution and shape are estimated by MC predictions.

Figure \ref{fig:2MBackground} shows the $m^{\text{reco}}_{\text{inv}}$ spectra including the sideband fit for the signal selection for the 2M and 1M1S categories, respectively. The expected signal shape for an ALP with $m_a = 0.5\,\GeV$ and $C_{a\gamma\gamma} = 0.01$ is also shown for illustration.

\begin{figure*}[thb]
\centering
\subfloat[]{
\includegraphics[width=0.48\textwidth]{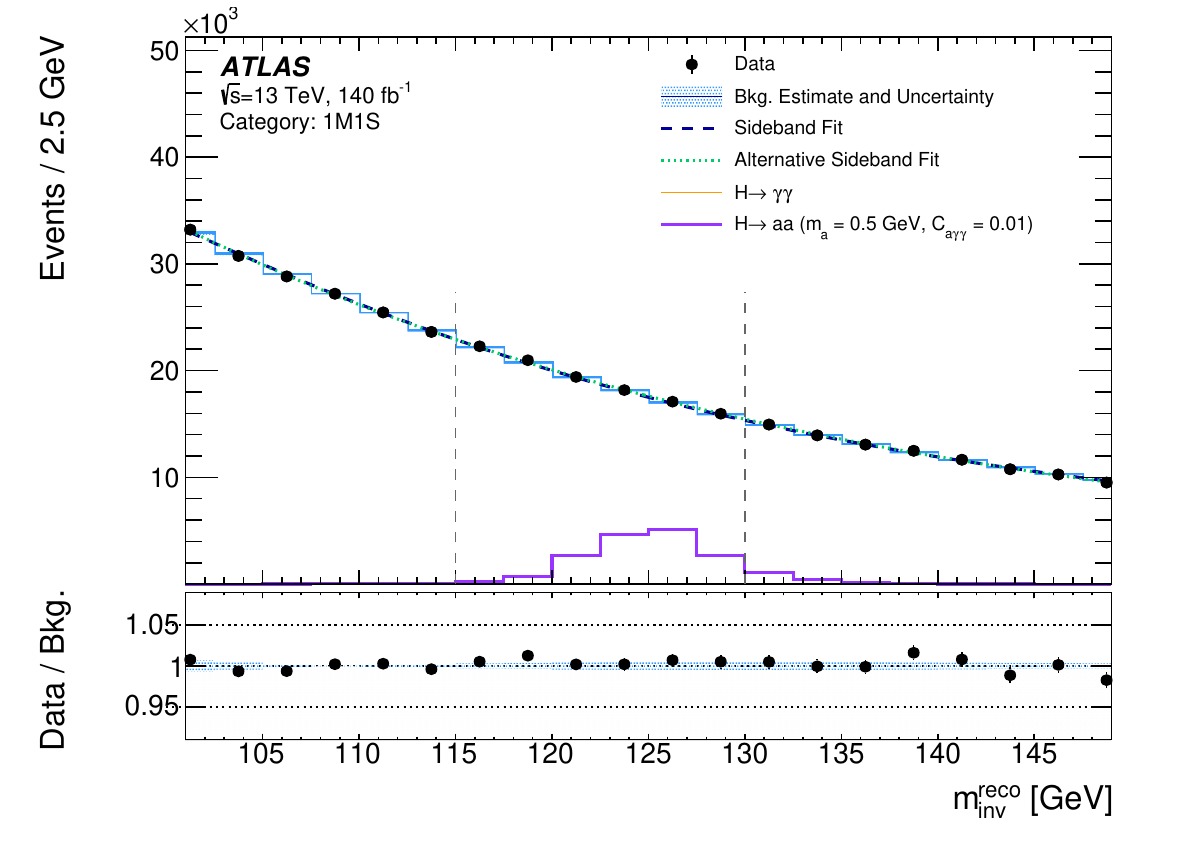}
} \hfill
\subfloat[]{
\includegraphics[width=0.48\textwidth]{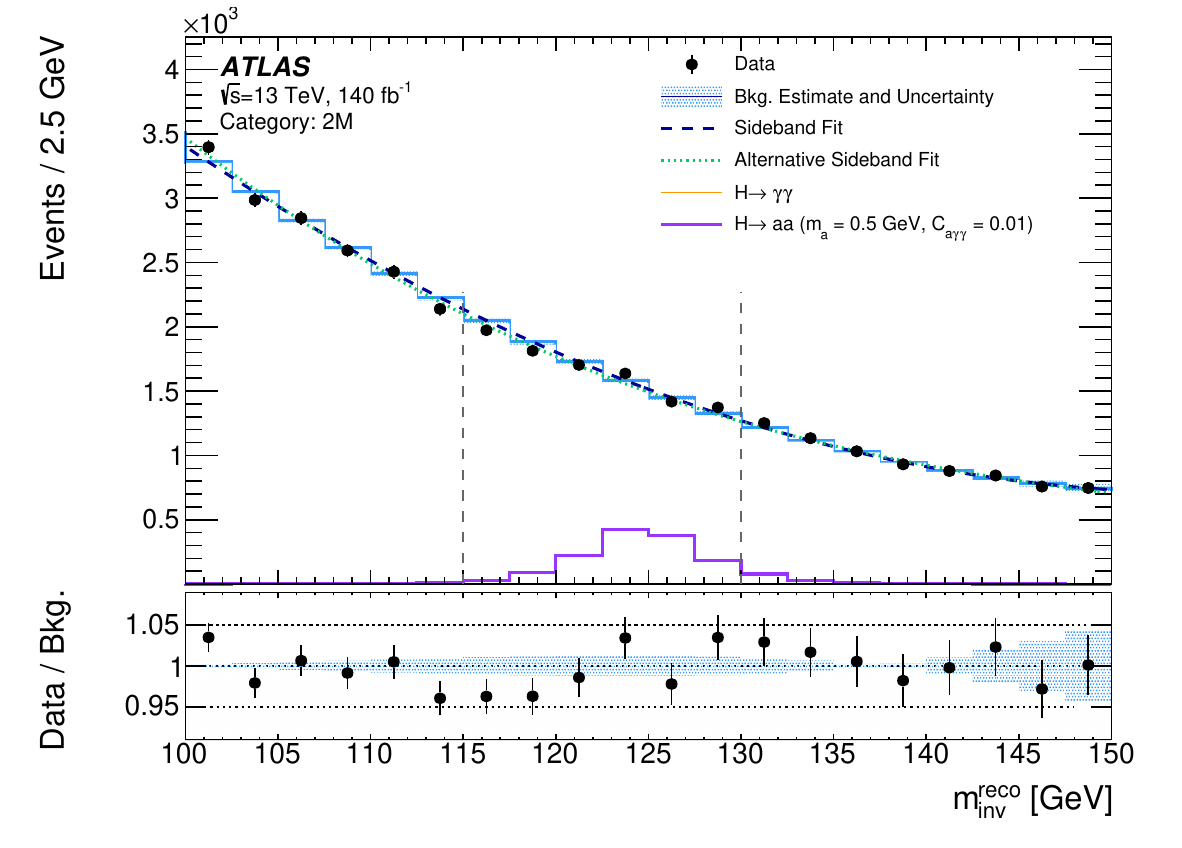}
} \hfill
\caption{$m^{\text{reco}}_{\text{inv}}$ distribution for the nominal signal selection for the (a) 1M1S and (b) 2M category. The nominal sideband fitting function is shown as the blue dashed line. The background, estimated from a fit in the side-band regions, and its systematic uncertainty
is shown as a blue histogram for both cases. The green dotted line shows the alternative fitting function used to estimate the spurious signal uncertainty (discussed in Section \ref{sec:systematics}). The expected signal shape for $m_a=0.5\,\GeV, C_{ayy}=0.01$ is also shown with an arbitrary normalization. The signal region selection on $m^{\text{reco}}_{\text{inv}}$ is indicated using vertical dashed lines. The contribution from $H\rightarrow\gamma\gamma$ is negligible and not visible in the figures. The lower panels show the data divided by the estimated continuum background, where the shaded area indicates the uncertainty on the background estimation. }
\label{fig:2MBackground}
\end{figure*}

\subsection{Three- and Four-Photon Final States in the Search for Long-Lived Axion-Like Particles\label{sec:background:3y}}

The background estimation in the long-lived ALPs searches also employs a sideband fit using the $m^{\text{reco}}_{\text{inv}}$ spectrum in the 3S and 4S categories.

Polynomials of third and second order serve as the nominal background fitting functions for the 3S and 4S categories, respectively, where the fits are carried out in the range of 80~\GeV\ to 150~\GeV\ and 105~\GeV\ to 145~\GeV, excluding the signal region.
First, the suitability of the sideband functions for background estimation in both categories is tested on three- and four-photon continuum MC samples.
Next, the sideband functions and corresponding background estimates are validated using an orthogonal set of data events in which the requirement on the reconstructed ALP mass is inverted.

This inverted sample can be used as a validation region, since the shape of background events should not change with a different choice of ALP mass apart from minor kinematic changes in the $m^{\text{reco}}_{\text{inv}}$ distribution.
The multi-photon MC samples are used to correct for this kinematic effect.

The chosen fitting functions yield a $\chi^2/{\text{ndf}}$ close to unity in all validation regions and the estimated background using these validation regions is consistent with the observed numbers of background events from the signal region sidebands.
Systematic uncertainties due to the choice of the background function and the fitting range are discussed in Section~\ref{sec:systematics}.
Figure \ref{fig:4SLongLived} depicts the $m^{\text{reco}}_{\text{inv}}$ spectrum for various ALP mass searches, and shows the sideband fitting functions, the estimated background in the signal region, and the expected signal shape for two $C_{a\gamma\gamma}$ coupling parameters.

\begin{figure*}[thb]
\centering
\subfloat[]{
\includegraphics[width=0.48\textwidth]{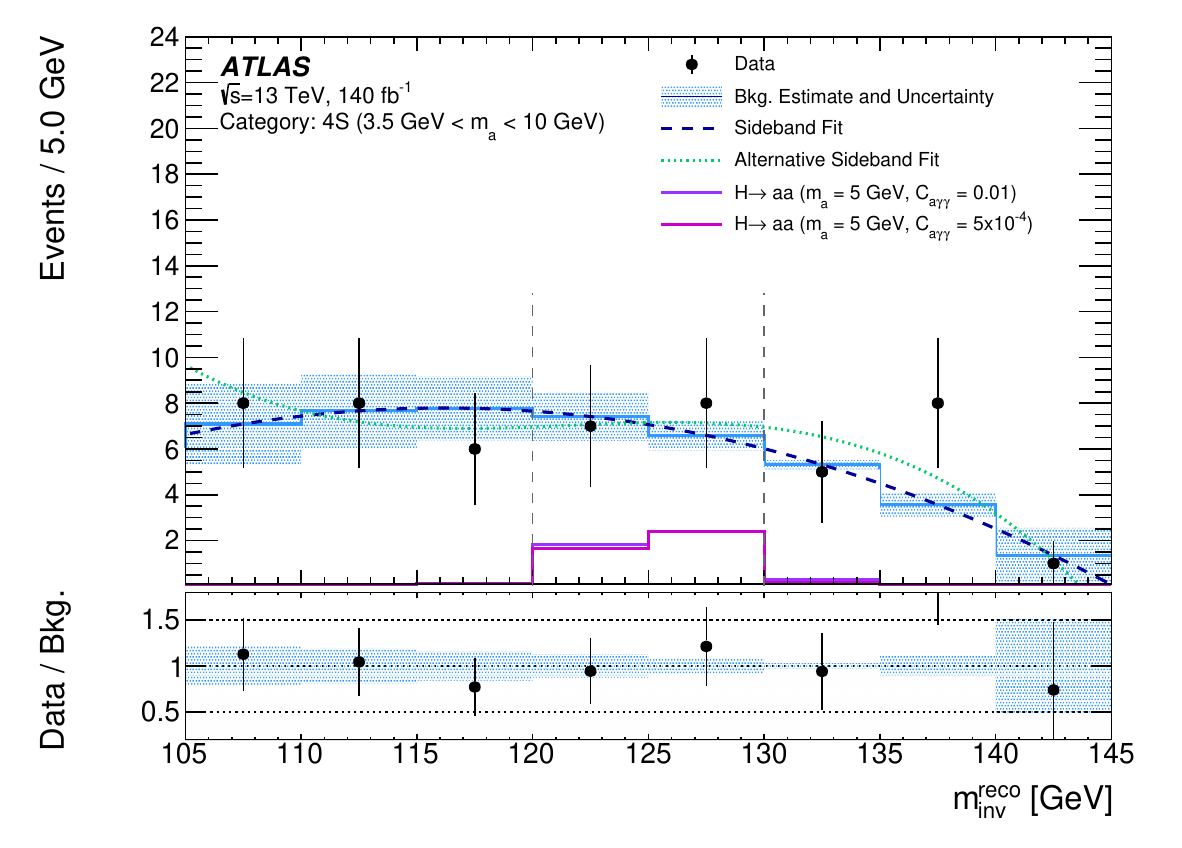}
} \hfill
\subfloat[]{
\includegraphics[width=0.48\textwidth]{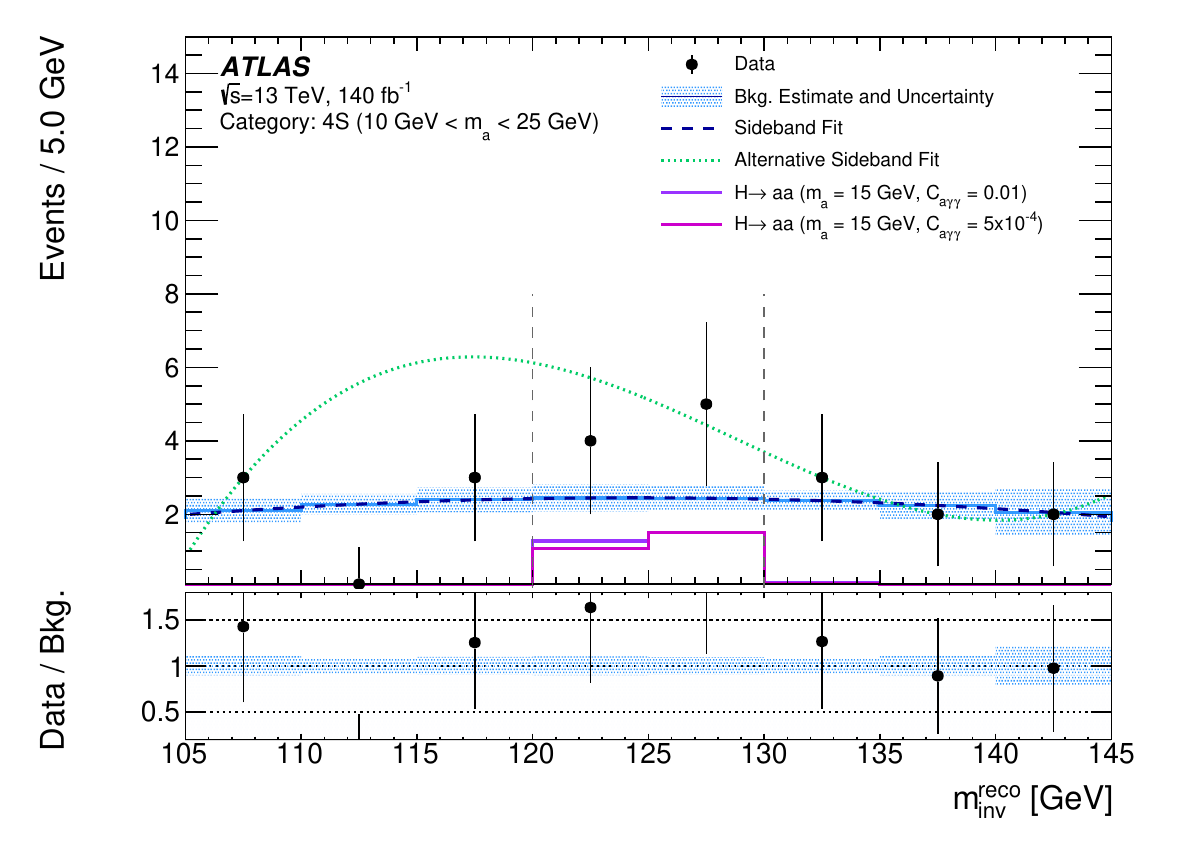}
} \hfill
\subfloat[]{
\includegraphics[width=0.48\textwidth]{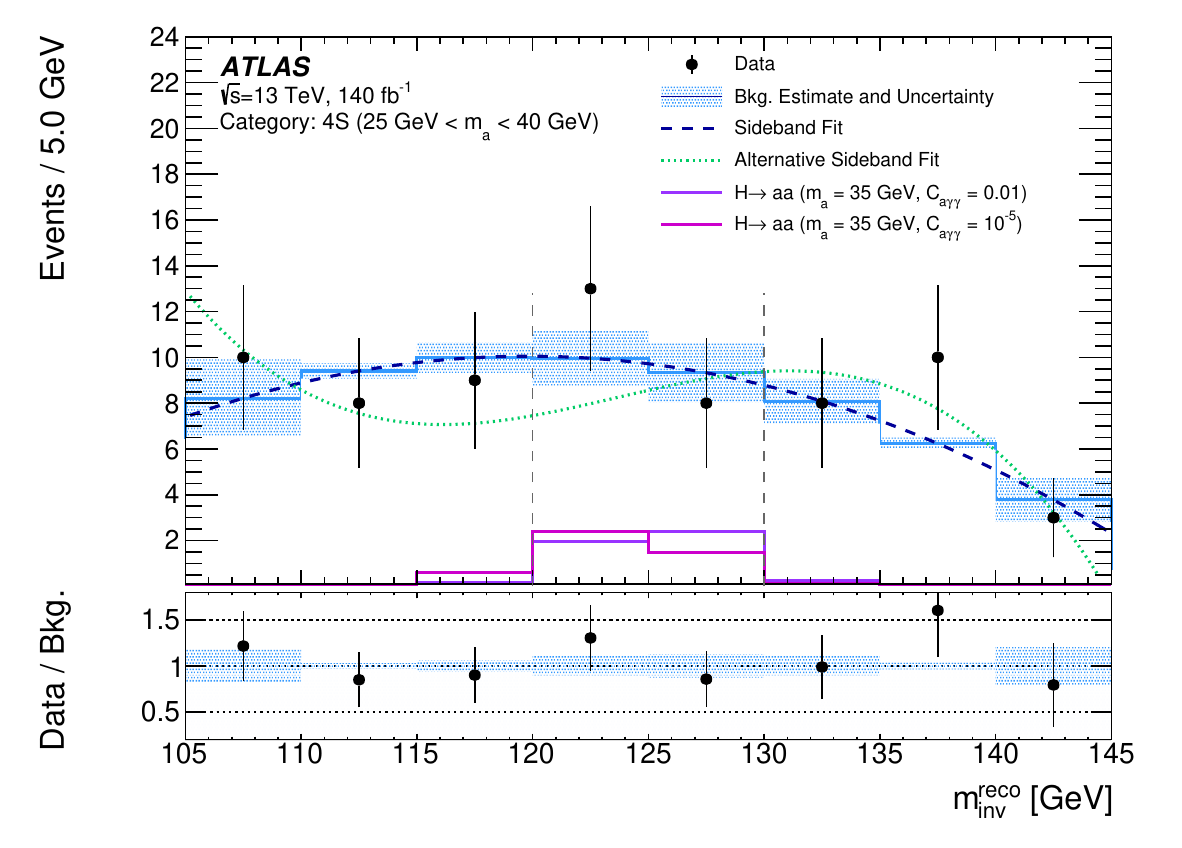}
} \hfill
\subfloat[]{
\includegraphics[width=0.48\textwidth]{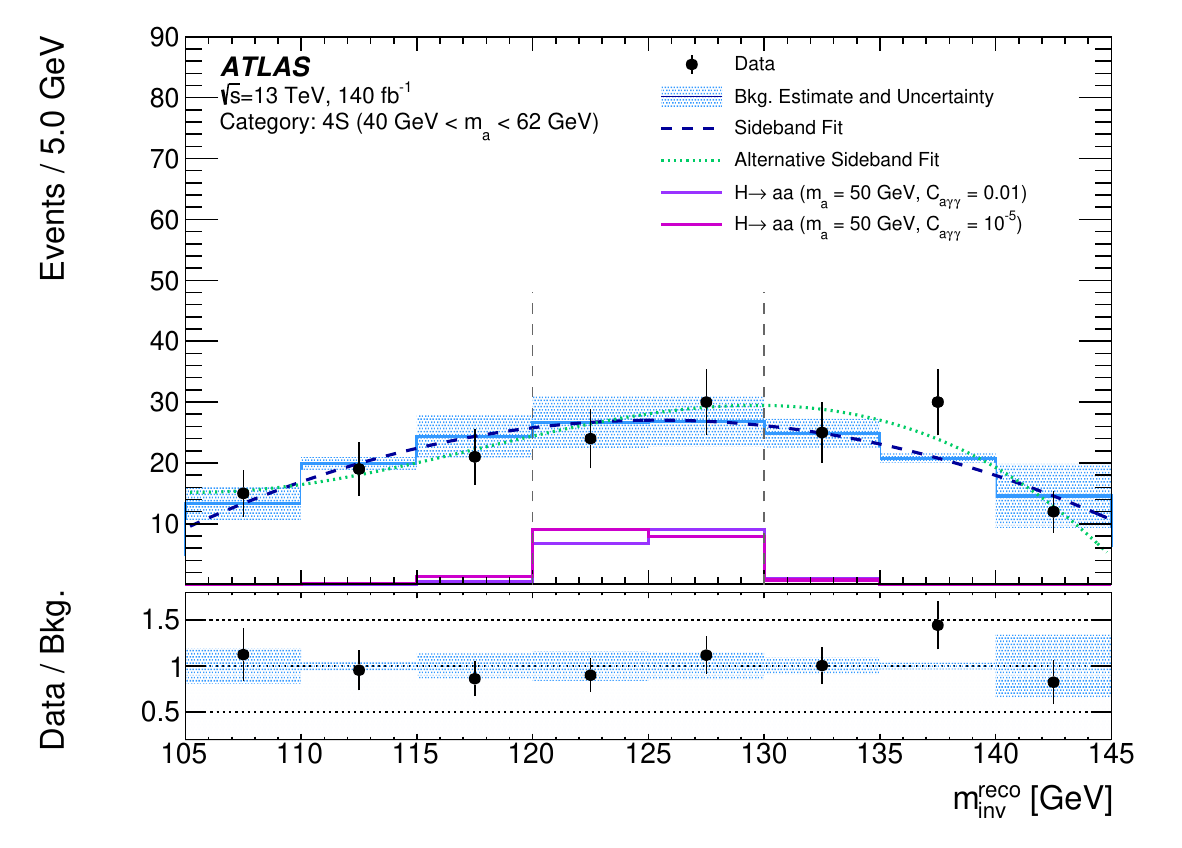}
} \hfill
\caption{$m^{\text{reco}}_{\text{inv}}$ distribution for the nominal signal selection for the 4S category. The nominal sideband fitting function is shown as the blue dashed line.
The background, estimated from a fit in the side-band regions, and its systematic variation (obtained from a fit with reduced range) is shown as the blue histogram.
The green dotted line shows the alternative fitting function which is used to estimate the spurious signal uncertainty (discussed in Sec. \ref{sec:systematics}). The four subfigures show different ALP mass ranges: (a) $3.5\,\GeV<m_a<10\,\GeV$, (b) $10\,\GeV<m_a<25\,\GeV$, (c) $25\,\GeV<m_a<40\,\GeV$, (d) $40\,\GeV<m_a<62\,\GeV$. The signal region selection on $m_{a}^{\text{reco}}$ is applied while the signal region selection on $m^{\text{reco}}_{\text{inv}}$ is indicated as dashed lines. The expected signal shapes for $m_a=5\,\GeV, C_{ayy}=0.01$; $m_a=5\,\GeV, C_{ayy}=5\cdot10^{-4}$; $m_a=15\,\GeV, C_{ayy}=0.01$; $m_a=15\,\GeV, C_{ayy}=5\cdot10^{-4}$;  $m_a=35\,\GeV, C_{ayy}=0.01$; $m_a=35\,\GeV, C_{ayy}=10^{-5}$;  $m_a=50\,\GeV, C_{ayy}=0.01$; $m_a=50\,\GeV, C_{ayy}=10^{-5}$;is shown with arbitrary normalization. The lower panels show the data divided by the estimated continuum background, where the shaded area indicates the uncertainty on the background estimation. }
\label{fig:4SLongLived}
\end{figure*}

\subsection{Four-Photon Final States in the Search for Promptly Decaying Axion-Like Particles}

The number of selected events in the $4S_p$ category of the search for promptly decaying ALPs, also defined in the $m^{\text{reco}}_{\text{inv}}$ vs. $m^{\text{reco}}_{a}$ plane, is significantly lower than that in the analysis optimised for long-lived ALPs due to the stricter rejection of fake-photon signatures using more stringent selection criteria such as Tight photon identification.
To further suppress background contributions in the signal region, a tight selection around the $m_a$ model parameter, as discussed in Section \ref{sec:eventobjects}, is imposed.
The size of the signal regions for each ALP mass are shown in Table \ref{tab:SignalRegions}.
The background in the signal region of the search for promptly decaying ALPs
is estimated by counting the events around the signal region in the
$m^{\text{reco}}_{\text{inv}}$-$m^{\text{reco}}_{a}$ plane, extending the signal region by $\pm5\,\gev$ in the $m^{\text{reco}}_{\text{inv}}$  dimension and by $1.5$ times the signal region width in the $m^{\text{reco}}_{a}$ dimension.
Due to the low statistics, the total number of background events in the shaded sideband area is scaled by the ratio of the areas of the signal region to the sideband region to estimate the number of background events in the signal region, as illustrated in Figure \ref{fig:promptbkg}.
This is equivalent to assuming a flat background distribution in the plane.

As an alternative background estimate, the size of the control region is taken to be 2.5 times the signal region instead of the 1.5 times used in the nominal background estimate.
The difference between the nominal and the alternative background estimates is used as a systematic uncertainty.
Figure \ref{fig:promptbkg}
shows distributions in the $m^{\text{reco}}_{\text{inv}}$-$m^{\text{reco}}_{a}$ plane for events selected in the search for promptly decaying ALPs, with the sideband regions shown for the search parameters $m_a=10$ GeV and $m_a=40$ GeV. The validity of this procedure was tested using simulated multi-photon samples
confirming that the associated statistical and systematic uncertainties cover potential shape differences.

\begin{figure*}[thb]
\centering
\subfloat[]{
\includegraphics[width=0.48\textwidth]{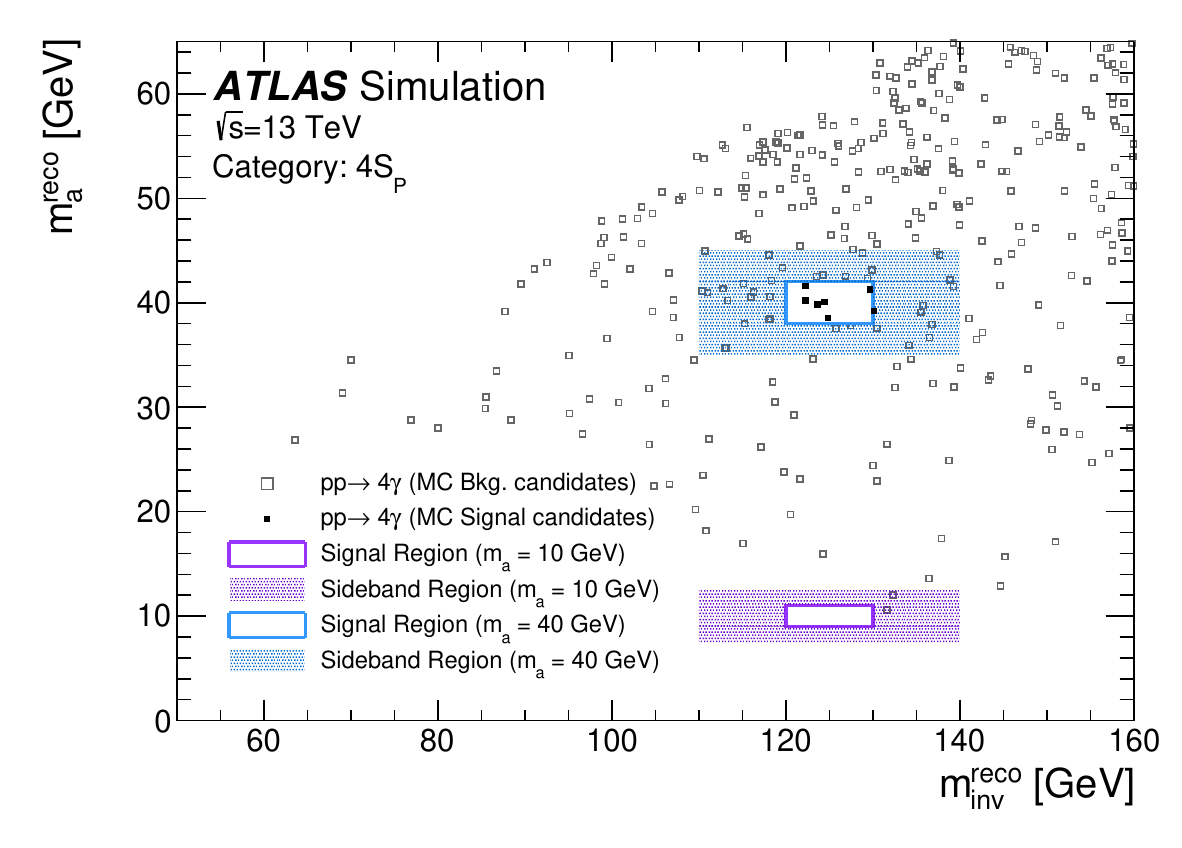}
} \hfill
\subfloat[]{
\includegraphics[width=0.48\textwidth]{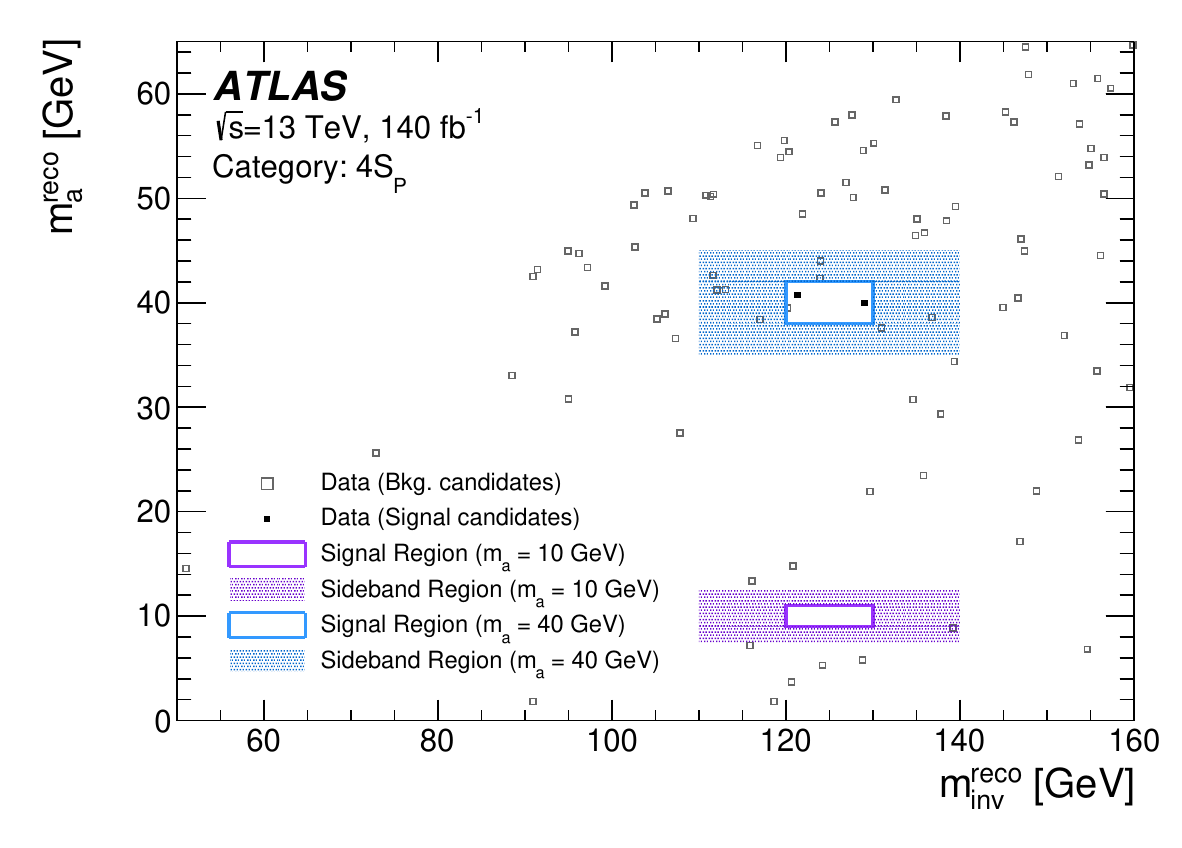}
} \hfill
\caption{$m^{\text{reco}}_{\text{inv}}$ vs. $m^{\text{reco}}_{a}$ for the $4S_p$ categories in the search for promptly decaying ALPs, for (a) simulated $pp \rightarrow 4\gamma$ sample and (b) for data. The signal (sideband) regions are indicated by solid lines (shaded areas) for the searches for ALPs with masses of 10~GeV and 40~GeV. Events within the signal region are shown with filled markers, those outside with open markers. }
\label{fig:promptbkg}
\end{figure*}

\subsection{Background Estimate Summary} %
Table \ref{tab:cutflow} summarizes the data and the expected background contribution in the signal region for the different categories. The acceptance times selection efficiency of a signal event in any of the categories is largely dependent on the ALP mass and coupling parameters under investigation. It ranges from $9\%$ to $23\%$ for low ALP masses ($m_a<5$ GeV) and large couplings ($C_{a\gamma\gamma}=1$) and is around 13\% for large ALP masses ($m_a \approx 60$ GeV) at all studied couplings.
The largest fraction of events are already cut by the trigger due to the relatively high trigger thresholds.
A future reduction of the trigger thresholds could significantly improve the analysis acceptance.

\begin{table}[ht]
\small
\begin{center}
\begin{tabular}{l | cc || l | cc}
\hline
\multicolumn{3}{c||}{Long-lived ALP Search} & \multicolumn{3}{c}{Prompt ALP Search} \\
\hline
Category     & Events    & Background & Category    & Events    & Background \\
$C_{a\gamma\gamma} < 0.1$  &  & Estimate & $0.1 < C_{a\gamma\gamma} < 1$  &  & Estimate \\
\hline\hline
2M                  & 9917     &   $10000 \pm 200$      & $4\texttt{S}_p$ ($m_a$ = 5 GeV) 	& 2 & $0.35\pm0.27$ \\
\hline
1M1S                &  113875    &    $113100 \pm 700$    & $4\texttt{S}_p$ ($m_a$ = 10 GeV) 	& 0 & $0.24\pm0.18$ \\
\hline
2S                  &   632484   &    $634000 \pm 1500$    & $4\texttt{S}_p$ ($m_a$ = 15 GeV) & 1 & $0.24\pm0.28$ \\
\hline
3S ($3.5\,\text{GeV} \leq m_a< 10\,\text{GeV}$) &   4782   &    $4800  \pm 100$     & $4\texttt{S}_p$ ($m_a$ = 20 GeV) 	& 0 & $0.24\pm0.28$ \\
\hline
3S ($10\,\text{GeV} \leq m_a< 25\,\text{GeV}$) &  3166    &   $3200  \pm 100$      & $4\texttt{S}_p$ ($m_a$ = 25 GeV) 	& 0 & $0.24\pm0.28$ \\
\hline
3S ($25\,\text{GeV} \leq m_a<40\,\text{GeV}$) &  3843    &   $3900 \pm 150$    & $4\texttt{S}_p$ ($m_a$ = 30 GeV) 	    & 0 & $0.60\pm0.27$ \\
\hline
3S ($40\,\text{GeV} \leq m_a\leq 62\,\text{GeV}$) &  5389     &   $5600 \pm 90$      & $4\texttt{S}_p$ ($m_a$ = 35 GeV) 	& 0 & $1.31\pm0.76$ \\
\hline
4S ($3.5\,\text{GeV} \leq m_a<10\,\text{GeV}$) &   15   &   $14 \pm 3$       & $4\texttt{S}_p$ ($m_a$ = 40 GeV) 		    & 2 & $1.55\pm0.94$ \\
\hline
4S ($10\,\text{GeV} \leq m_a<25\,\text{GeV}$) &   9   &   $4 \pm 2$       & $4\texttt{S}_p$ ($m_a$ = 45 GeV) 			& 2 & $3.23\pm0.95$ \\
\hline
4S ($25\,\text{GeV} \leq m_a<40\,\text{GeV}$) &  21    &   $19 \pm 3$      & $4\texttt{S}_p$ ($m_a$ = 55 GeV) 		    & 7 & $2.84\pm1.51$ \\
\hline
4S ($40\,\text{GeV} \leq m_a\leq62\,\text{GeV}$) &   54    &   $54  \pm 6$       & $4\texttt{S}_p$ ($m_a$ = 62 GeV) 		& 4 & $4.52\pm1.50$ \\
\hline
\end{tabular}
\end{center}
\caption{Overview of the number of observed events in the search for long-lived ALPs (left) and selected mass points from the search for prompt ALPs (right) in comparison to the expected number of background events. The uncertainty on the background estimate includes statistical and systematic uncertainties as described in Section \ref{sec:systematics}.}
\label{tab:cutflow}
\end{table}


\section{Systematic Uncertainties\label{sec:systematics}}

The systematic uncertainties are assessed below, and their impact on the results is discussed in Section~\ref{sec:results}.
First the general experimental uncertainties are discussed, where special attention is given to the uncertainties arising from the displaced decay of long lived ALPs. Then the uncertainties impacting the background estimation are detailed followed by a discussion of the relevant theoretical uncertainties.

The experimental systematic uncertainty ranges, depending on the category and the hypothesized ALP mass and coupling, from 6.5\% to 18\% for most categories, with the exception of the 4S category where the uncertainty rises to 40\% for masses $m_a < 15\,\GeV$ and small couplings $C_{a\gamma\gamma}$ due to low statistics and large contributions from the photon identification uncertainties. The theoretical uncertainty is around 6\% for all categories.

\subsection{General Experimental Uncertainties}
The following general experimental systematics are applied to the signal model.

The uncertainty in the combined 2015--2018 integrated luminosity is 0.83\% \cite{ATLAS:2022hro}, obtained using the LUCID-2 detector \cite{Avoni:2018iuv} for the primary luminosity measurements, complemented by measurements using the inner detector and calorimeters.

To evaluate any impact on the expected signal yield due to imperfect modelling of pile-up, the average number of pile-up interactions is varied in the simulation. The corresponding uncertainty is below 1\%.

The trigger efficiency used to select events is evaluated in simulation and data using a bootstrap method and radiative $Z$-boson decays~\cite{TRIG-2018-05}. The difference between data and simulation, which ranges from 2\% to 3\%, is taken as a systematic uncertainty.

The systematic uncertainties from the standard photon identification and isolation efficiencies are estimated following the prescriptions in Ref. \cite{EGAM-2018-01}.
They affect the di-photon selection efficiency and are evaluated by varying the correction factors for photon selection efficiencies in simulation by their corresponding uncertainties.
The experimental uncertainties in the photon energy scale and resolution are obtained as described in Ref. \cite{EGAM-2018-01}.
These variations produce uncertainties below 3\% on the expected number of events in the signal region in the search for promptly decaying ALPs.

The uncertainties of the NN-based classifiers are estimated by comparing their identification performance using $Z\rightarrow ee$ events in simulation and data, where the electron shower-shape variables are used as the network input variables.
Very good agreement of the network output between data and simulation is observed.
The residual differences are fully propagated as uncertainties on the expected signal yields and produce normalisation uncertainties of up to 15\% in the 2M and 1M1S categories, respectively.

\subsubsection{Uncertainties due to displaced ALP decays}
The uncertainties related to photon identification and energy reconstruction for photons produced with displaced vertices -- i.e., those arising from long-lived ALP decays -- are estimated by studying the decays of long-lived hadrons, mainly kaons with transverse momentum $\pt > 10\,\GeV$, which can be reconstructed as displaced tracks in the ATLAS tracking system.
The daughter tracks from these decays, which originate from a displaced vertex,
can be matched to reconstructed clusters in the electromagnetic calorimeter.
A comparison of the shower shapes predicted by simulations to those from data for signatures with displaced vertices can then be used to estimate systematic uncertainties in photon reconstruction.
The MC prediction of the shower shapes of hadronic particles also relies on the correct description of particle multiplicities and energies.
To correct for any mismatch of particle composition in data and simulation,
scale factors are derived from the differences between data and MC predictions for shower-shape variables of tracks close to the primary vertex
($z_0<20$ mm and $d_0<1$ mm, where $z_0$ and $d_0$ are the longitudinal distance from the IP and the impact parameter, respectively).
These deviations are taken as a nominal bias and are used to correct tracks originating from a distance between $20\,\text{mm}<z_0<500\,\text{mm}$ and $1\,\text{mm}<d_0<80\,\text{mm}$ from the primary vertex (medium regime),
and tracks with an origin further than $z_0 > 500$ mm and $d_0 > 80$ mm from the interaction point (far regime).
These systematic uncertainties are used additionally for all photons stemming from a displaced decay.
The systematic uncertainty from the modelling of the NN classifier that discriminates real photons from fakes due to displaced photon vertices is $3\%$.
The corresponding uncertainties from the modelling of the photon identification and the second NN, which discriminates between merged and resolved photons, ranges from $4\%$ to $23\%$, depending on the displacement.

Long-lived hadrons are also utilized to estimate systematic uncertainties in energy reconstruction. The observed differences in the momentum over energy ratio between the prompt, medium and far regimes are found to be negligible compared to the nominal energy reconstruction uncertainty.

Systematic uncertainties in identifying the correct ALP pairing are mainly caused by variations in the photon energy scale corrections within their uncertainties. The final impact on the number of reconstructed events in a particular ALP mass signal region is less than 5\% and hence negligible.

\subsection{Uncertainties on the Background Estimation}
The continuum background processes are estimated from data and are subject to uncertainties related to the potential bias arising from the selected background model, as detailed in Section~\ref{sec:background}.
The nominal background estimate in each bin is calculated using the nominal fitting functions, which have been fitted in the sideband regions.
The shape uncertainty of the background is estimated by employing the same nominal fitting functions, but fitting them in a sideband region whose width is varied by 5~\GeV, corresponding to a 25\% to 100\% change in the fit range, depending on the category, allowing for a large variability of the background shape.
The spurious signal bias or background model bias is assessed as an additional uncertainty on the total number of signal events in each category.
This bias is estimated by generating pseudo-data using a modified background model and performing the full signal-plus-background fit (see Section~\ref{sec:results}) on these pseudo-data~\cite{ATL-PHYS-PUB-2020-028}.
The alternative background model for the spurious signal bias estimate is based on a second-order polynomial for the 2M, 1M1S, 3S, and 4S categories, while the function $f_{\text{sys}} = N_0 \exp(p\cdot x) + N_1 + a\cdot x^2 + b\cdot x$ is used for the 2S category.
The estimated signal-strength in these pseudo-data is considered as an additional systematic uncertainty in the final signal-strength estimation.
The largest impact is found in the 4S category shifting the branching ratio limit by $10^{-6}$.

\subsection{Impact of Theory Uncertainties}
To estimate the effects of scale uncertainties arising from missing higher-order corrections in the theoretical calculations, the factorisation and renormalisation scales are varied up and down by a factor of two from their nominal values. The cross-section is then recalculated for each case, and the largest deviation from the nominal cross-section is taken as the uncertainty. The uncertainties on the SM Higgs ggF production cross-section due to the choice of renormalisation scheme and top-quark mass, as well as their combination with those from factorisation and renormalisation scale variations, are based on Ref.~\cite{Baglio:2020wgt}.
The uncertainties in the cross-sections, which include the effects of uncertainties on the PDF and the strong coupling constant $\alpha_s$, and the uncertainties in the $H \rightarrow \gamma\gamma $ branching fractions, are taken from Ref.~\cite{deFlorian:2016spz} to be 5.7\%. It is found that further uncertainties on the Higgs boson signal prediction are negligible.

The uncertainties on the Higgs boson production cross-section
enter when calculating the limit on the branching ratio of the signal process on all Higgs boson decays. The values for the uncertainties are taken from Ref.~\cite{deFlorian:2016spz}.


\section{Results \label{sec:results}}

The statistical analysis is carried out using the PyHF framework~\cite{Heinrich:2021gyp,pyhf}.
In the long-lived ALP searches, the analysis results are obtained by performing a simultaneous maximum-likelihood fit to the $m^{\text{reco}}_{\text{inv}}$ distribution over the range $100~\GeV$ to $150~\GeV$ for the two most sensitive categories for each ALP mass and coupling parameter.
The 2M and 1S1M categories are most sensitive for low ALP masses ($m_a<5$~\GeV), while for larger ALP masses, the 4S category dominates over the 3S category.
Including more than two categories in the fit does not significantly improve the sensitivity for any model parameter.
In the prompt ALP search, only the number of events in the signal region of the $4S_p$ category is used.
The analysis sensitivity is limited by the available data statistics for the $4S$ and $4S_p$ categories, while the systematic uncertainties dominate in other categories.
The likelihood function is defined as follows:

\begin{equation}
\label{eqn:likelihood}
\mathcal{L} = {\displaystyle\prod_{c}\left(\mathrm{Pois}(n^{}_c|N_c(\pmb{\theta}))\cdot{\displaystyle\prod_{i=1}^{n_\textrm{bins}^c} f_c(n_i,m^{\text{reco}, i}_\text{inv},\pmb{\theta})} \right)\cdot G( \pmb{\theta})}.
\end{equation}

Here, $n_{\textrm{bins}}^c$ is the number of bins in the $m^{\text{reco}}_{\text{inv}}$ distribution, $n_c$ is the observed number of events, and $N_c$ is the expected number of events for each category $c$.
For each bin $i$ in the $m^{\text{reco}}_{\text{inv}}$ distribution of category $c$, $f_c$ is the value of the probability density function (pdf) which is estimated from simulation, $n_i$ is the number of observed events in bin $i$, $\pmb{\theta}$ represents the nuisance parameters (NP) used to parametrize the effect of systematic uncertainties, and $G(\pmb{\theta})$ represents constraint pdfs for the nuisance parameters. All constraints correspond to Gaussian pdfs.
The expected number of events $N_c$ is defined as the sum of the expected yields from $H\to aa \to 4\gamma$ production processes ($N_{H\to aa}$), single Higgs-boson production ($N^{\mathrm{H\to\gamma\gamma}}_{\mathrm{bkg}}$), the non-resonant background ($N^{\mathrm{nonres}}_{\mathrm{bkg,c}}$), and the spurious signal uncertainty ($N_{\mathrm{SS,c}}$). It is defined as:

\begin{equation}
N_c( \pmb{\theta}) = \mu \cdot N_{H\to aa}( \pmb{\theta}_{H\to aa}^\mathrm{yield}) + N^{\mathrm{H\to\gamma\gamma}}_{\mathrm{bkg}}( \pmb{\theta}_{\mathrm{H\to\gamma\gamma}}^\mathrm{yield}) + N^{\mathrm{nonres}}_{\mathrm{bkg,c}}(\pmb{\theta}_\mathrm{nonres}^\mathrm{yield}) + N_{\mathrm{SS,c}}.
\end{equation}

Here, $\mu$ is the signal strength, and $ \pmb{\theta}^\mathrm{yield}$ represents the NPs affecting the event yield, as described in Section~\ref{sec:systematics}.
Correlation of the nuisance parameters across different signal and background components, and categories, is taken into account.
The normalisation parameter for the $H\rightarrow \gamma\gamma$ production rate, $\pmb{\theta}_{\mathrm{H\to\gamma\gamma}}^\mathrm{yield}$, is set to 1 and corresponds to the SM prediction for the $H\rightarrow \gamma\gamma$ cross-section. It is allowed to vary within its theoretical and experimental uncertainties.

The signal-plus-background hypothesis for the production of a Higgs boson that decays into ALPs is tested using the  profile-likelihood-ratio test statistic derived from Eq. \ref{eqn:likelihood}, and is parameterized with the signal-strength parameter $\mu$.
This parameter is defined as the ratio of the extracted signal events to the total number of signal events in the MC simulation.\footnote{Only $H\rightarrow aa \rightarrow 4\gamma$ events are simulated in the signal samples, i.e., the branching ratio for this process is unity in the simulation.}

Figure~\ref{fig:PostFitSignalYields} shows the distribution
of the number of estimated and observed events in the signal region of the most sensitive category for various ALP masses and coupling parameters for the
prompt and long-lived ALP searches.
Good agreement is observed between the estimated backgrounds and the data.
No significant pulls of the NP are observed after the fits.
The NP $\pmb{\theta}_{\mathrm{H\to\gamma\gamma}}^\mathrm{yield}$ is found to be consistent with 1,
corresponding to the expected SM Higgs boson production cross-section on the 2S category.

\begin{figure*}[tb]
\centering
\includegraphics[width=0.95\textwidth]{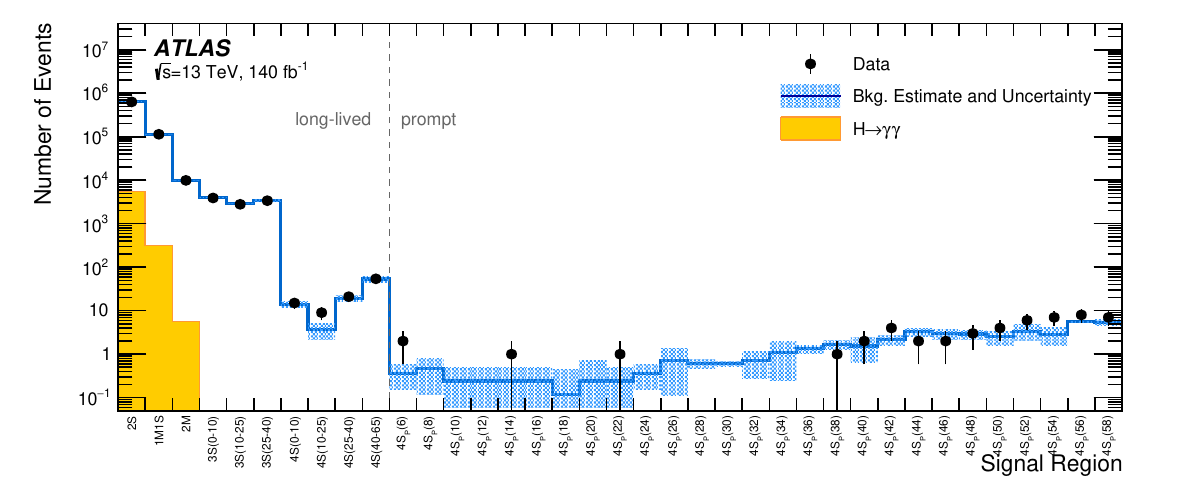}
\caption{The number of data and estimated background events in the signal region of the most sensitive categories.
The uncertainty in the background estimate is shown as shaded band. The left side shows the different categories of the long-lived ALP search, while the right side displays the $4S_p$ category of the prompt search for increasing mass hypotheses. The numbers in parentheses in the x-axis labels correspond to the probed ALP mass hypothesis in \GeV. The SM $H\rightarrow\gamma\gamma$ background is only sizeable in the first three bins, corresponding to the two-photon categories.}
\label{fig:PostFitSignalYields}
\end{figure*}

Upper limits, derived using the CL$\text{s}$ technique~\cite{Read:2002hq}, are set on $\mathcal{B}(H \rightarrow aa \rightarrow 4\gamma)$.
The branching ratio is obtained by dividing the fitted signal cross-section by the total Higgs-boson production cross-section of $55.6\, \text{pb}$ \cite{deFlorian:2016spz}.
The expected and observed limits as a function of $m_a$ from the search optimized for long-lived ALPs, i.e., $C_{a\gamma\gamma}<0.1$, are shown in Figure~\ref{fig:1DExclusionLongLived}
along with the performance on prompt decays for $C_{a\gamma\gamma}=1$.
For low masses and low couplings, the lifetime of the ALP gets significantly larger and most of the ALPs decay outside the active detector area. Therefore there are no limits available for $C_{a\gamma\gamma} \le 5\times 10^{-4}$ and m$_a<10$ \gev.
Limits for prompt decays with $C_{a\gamma\gamma}=1$ are also shown.
A relatively uniform sensitivity is achieved for larger ALP masses, above $10~\GeV$ for $C_{a\gamma\gamma} \ge 5\times 10^{-4}$ and above $25\,\GeV$ for $C_{a\gamma\gamma} = 10^{-5}$.
The sensitivity for low ALP masses decreases with smaller coupling values, as more ALP decays happen outside the sensitive detector volume.
The largest differences between expected and observed limits are found in long-lived ALP searches for masses between 10 and 25~\GeV\ at 1.5 $\sigma$.
It should be noted that the background estimation is the same for all couplings in this mass region. Hence correlated behaviour is expected for all relevant couplings.
The limits for the couplings $C_{a\gamma\gamma}=1$ %
and $C_{a\gamma\gamma}=0.01$ %
as a function of $m_a$ are shown in Figure~\ref{fig:1DExclusionLongLived_split}, separated for low ALP masses ($m_a <5\,\GeV$) and higher ALP masses. The upper limits at 95\% CL on $\mathcal{B}(H\rightarrow aa\rightarrow 4\gamma)$ range from $10^{-4}$ to  $3 \times 10^{-2}$ for low ALP masses and from $2 \times 10^{-5}$ to  $2 \times 10^{-4}$ for higher ALP masses.
The observed limits are compatible with the expected limits.
The loss in sensitivity around $3$~\GeV\ is due to the transition between the merged and resolved photon categories, where the former have significantly larger background contributions.

The limits on $\mathcal{B}(H \rightarrow aa \rightarrow 4\gamma)$ derived from the $4S_p$ category assuming promptly decaying ALPs are shown in Figure \ref{fig:1DExclusionPrompt}. The derived limit is mostly flat at $\mathcal{B} < 2\times 10^{-5}$ in the mass range $10\,\GeV < m_a < 62\,\GeV$.
For the long-lived searches it is necessary to loosen the selection criteria to allow for displaced ALP decays. Therefore the background contributions are significantly larger and consequently the searches are less sensitive than the prompt searches.
The observed limits are consistent with the expected limits.

The limits on ALP masses with $m_a>15$~\GeV\ are about one order of magnitude more stringent than previous ATLAS analyses~\cite{ATLAS:2015rsn} using $8\,\TeV$ data,
and reach similar to slightly better sensitivity than previous analyses from CMS~\cite{CMS:2022xxa} using $132\,\ifb$ of $\sqrt{s} = 13\,\TeV$ data.
These are the first limits on ALPs with masses below $10\,\GeV$ from the ATLAS experiment, and are up to 40\% more stringent than previous results from CMS~\cite{CMS:2022fyt} using $136\,\ifb$ of $\sqrt{s} = 13\,\TeV$ data.
The limits on long-lived ALPs in anomalous Higgs boson decays are the first obtained by any experiment.

\begin{figure*}[thb]
\centering
\subfloat[]{
\includegraphics[width=0.48\textwidth]{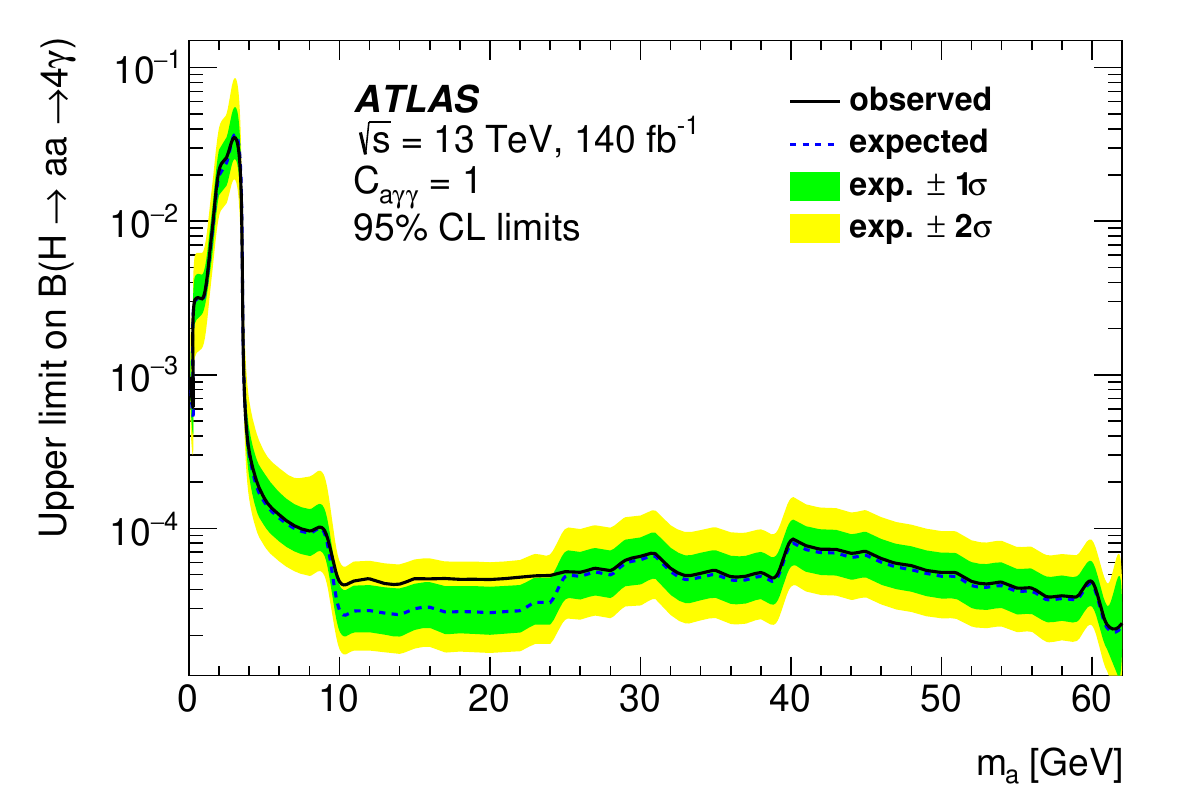}
}
\hfill
\subfloat[]{
\includegraphics[width=0.48\textwidth]{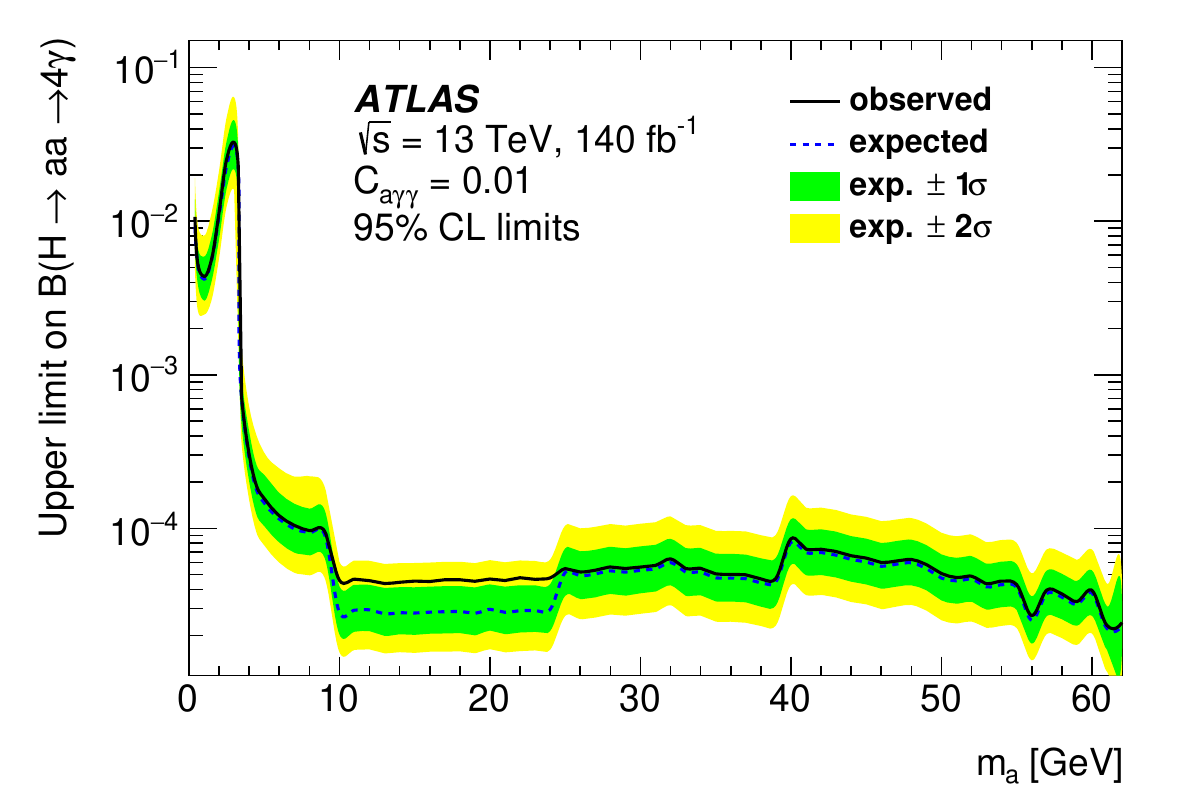}
}
\hfill
\subfloat[]{
\includegraphics[width=0.48\textwidth]{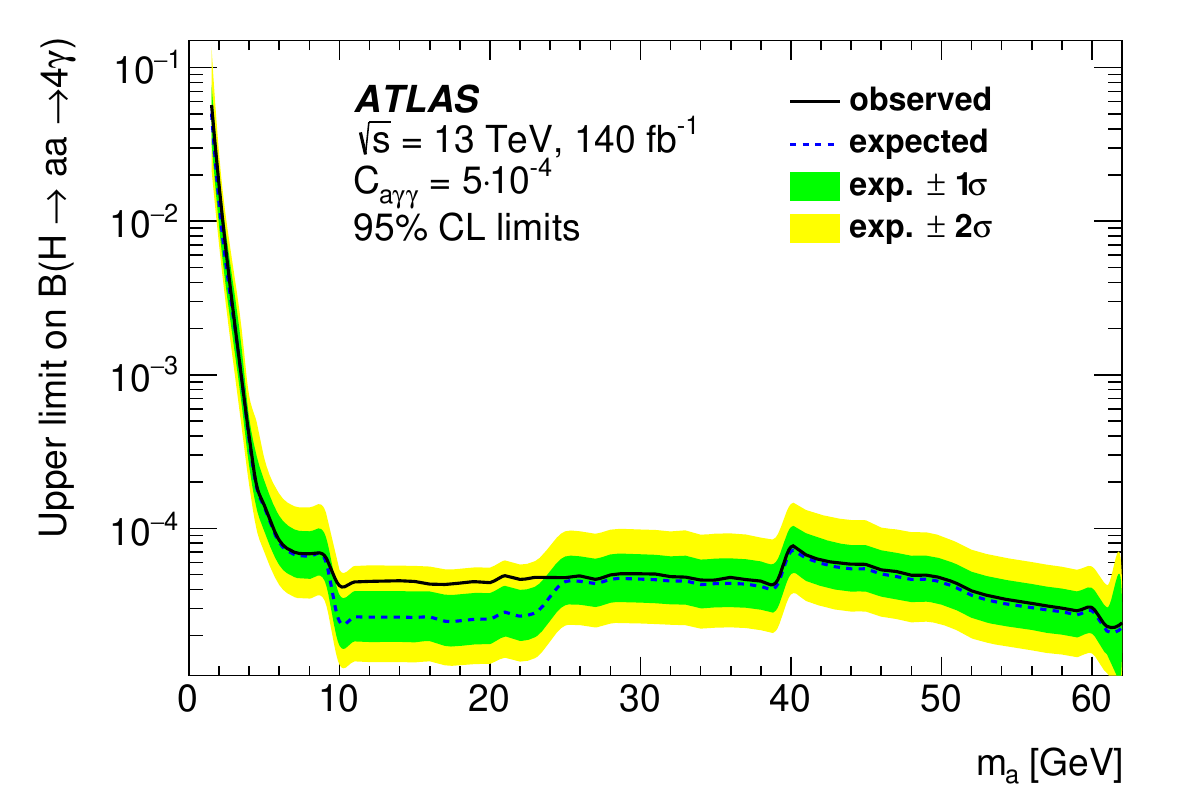}
}
\hfill
\subfloat[]{
\includegraphics[width=0.48\textwidth]{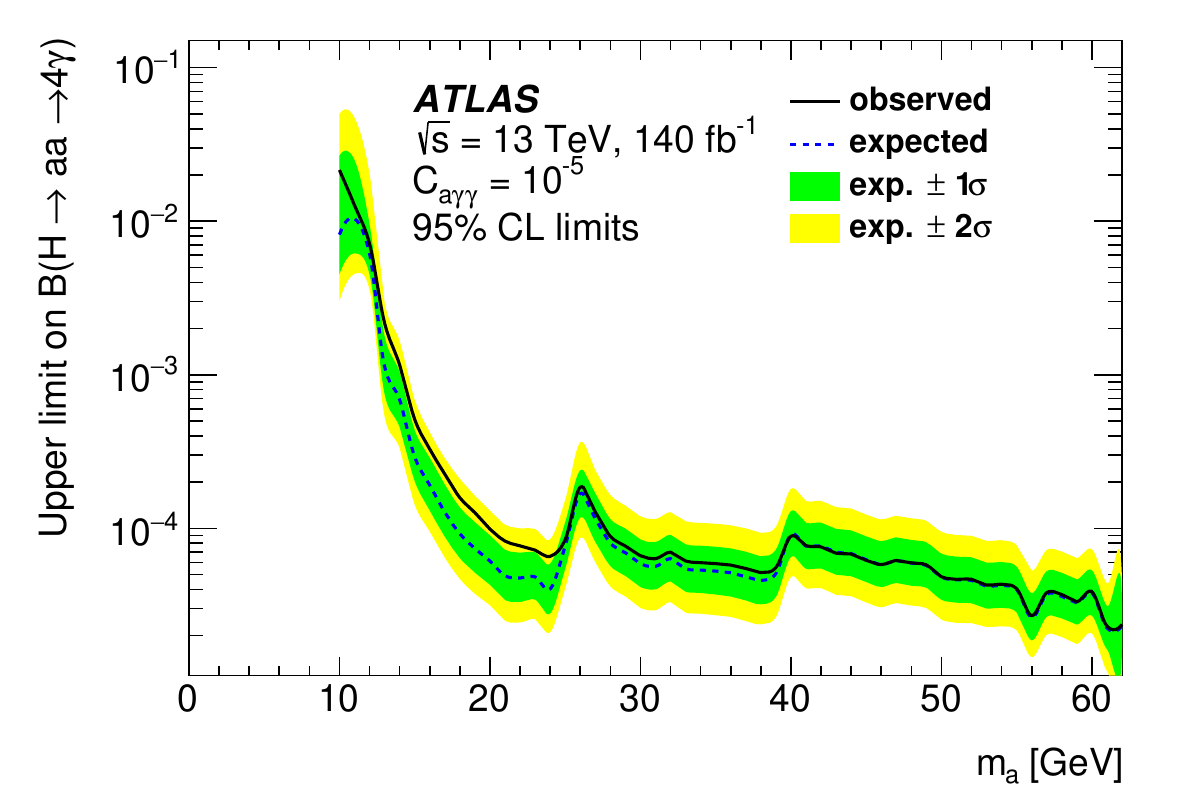}
}
\caption{Upper limits on $\mathcal{B}(H\rightarrow aa\rightarrow 4\gamma)$ at 95\% CL as a function of the axion mass and for different ALP-photon couplings, from (a) $C_{a\gamma\gamma}=1$ to (d) $C_{a\gamma\gamma}=10^{-5}$.}
\label{fig:1DExclusionLongLived}
\end{figure*}

\begin{figure*}[thb]
\centering
\subfloat[]{
\includegraphics[width=0.48\textwidth]{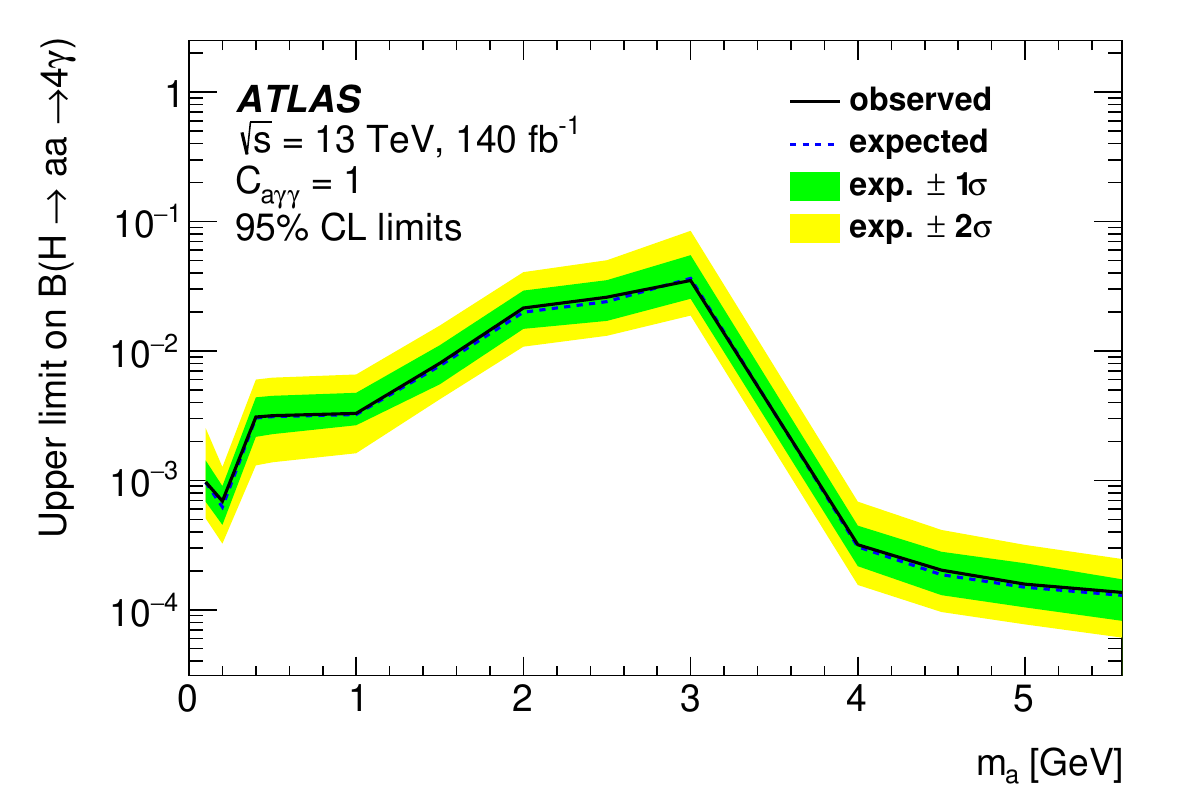}
}   \hfill
\subfloat[]{
\includegraphics[width=0.48\textwidth]{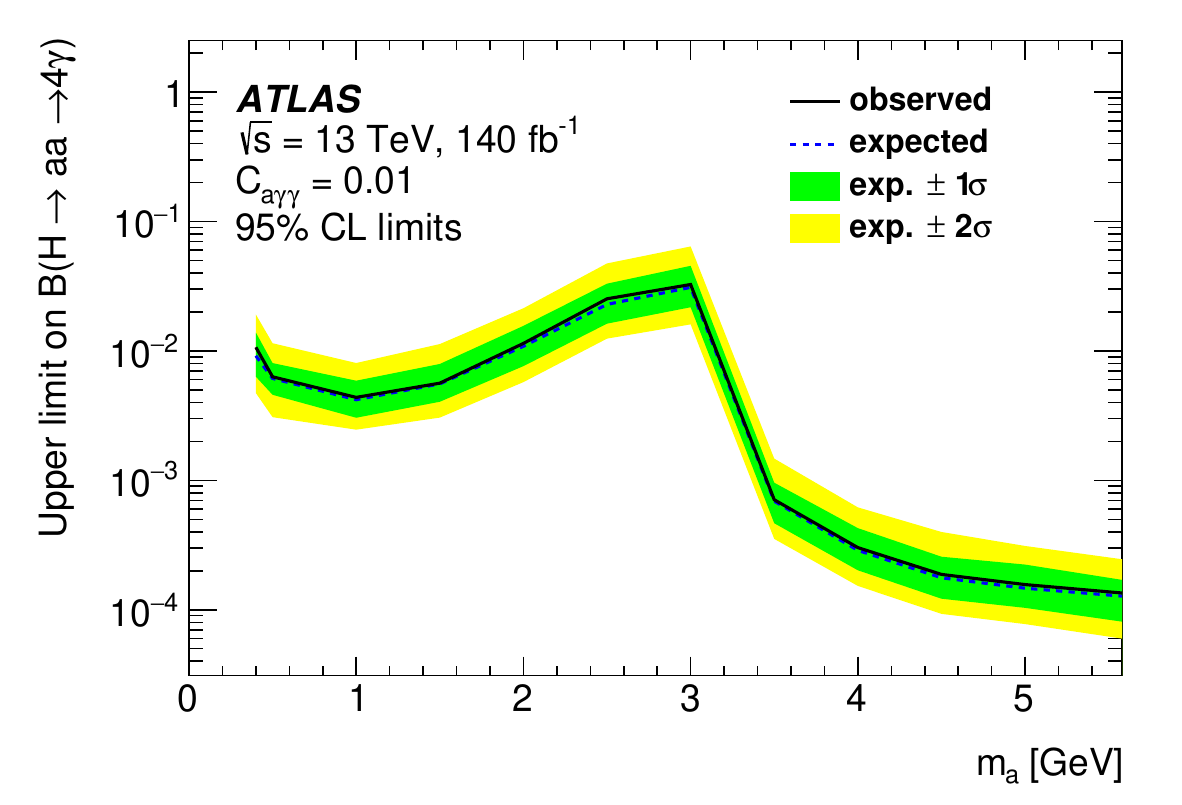}
}   \hfill
\subfloat[]{
\includegraphics[width=0.48\textwidth]{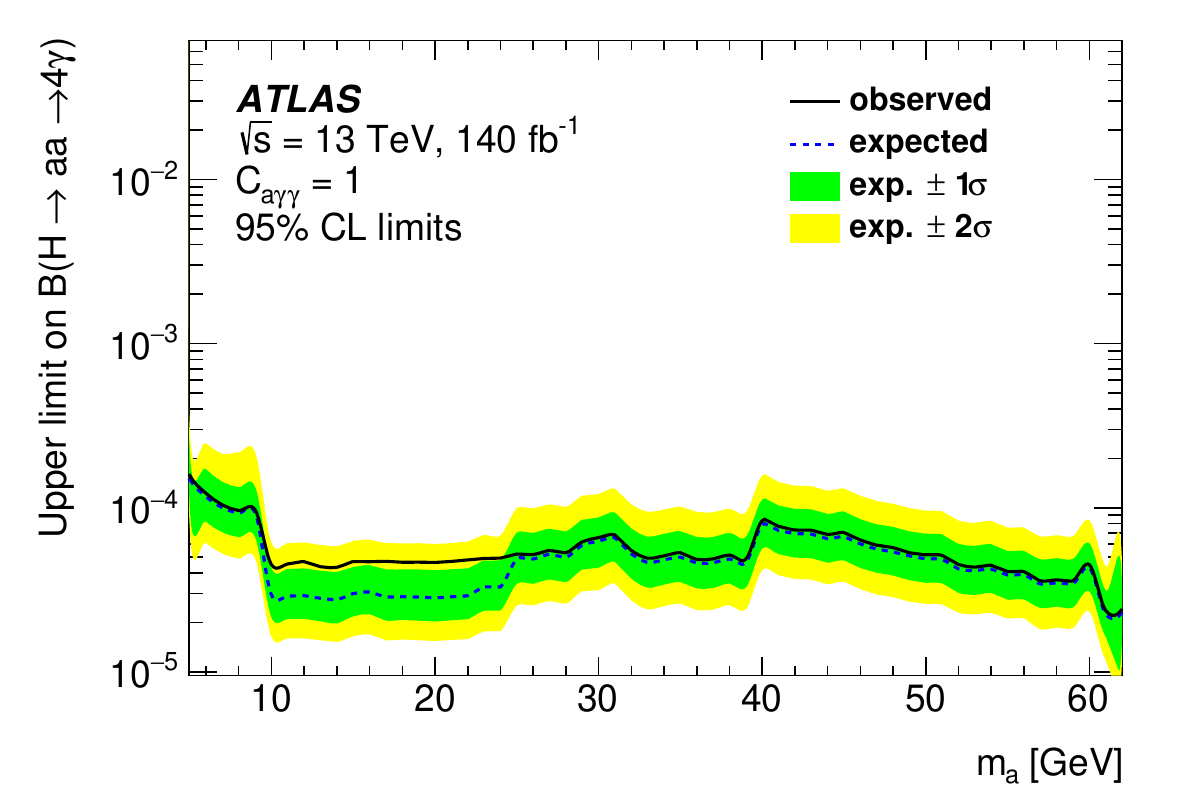}
}   \hfill
\subfloat[]{
\includegraphics[width=0.48\textwidth]{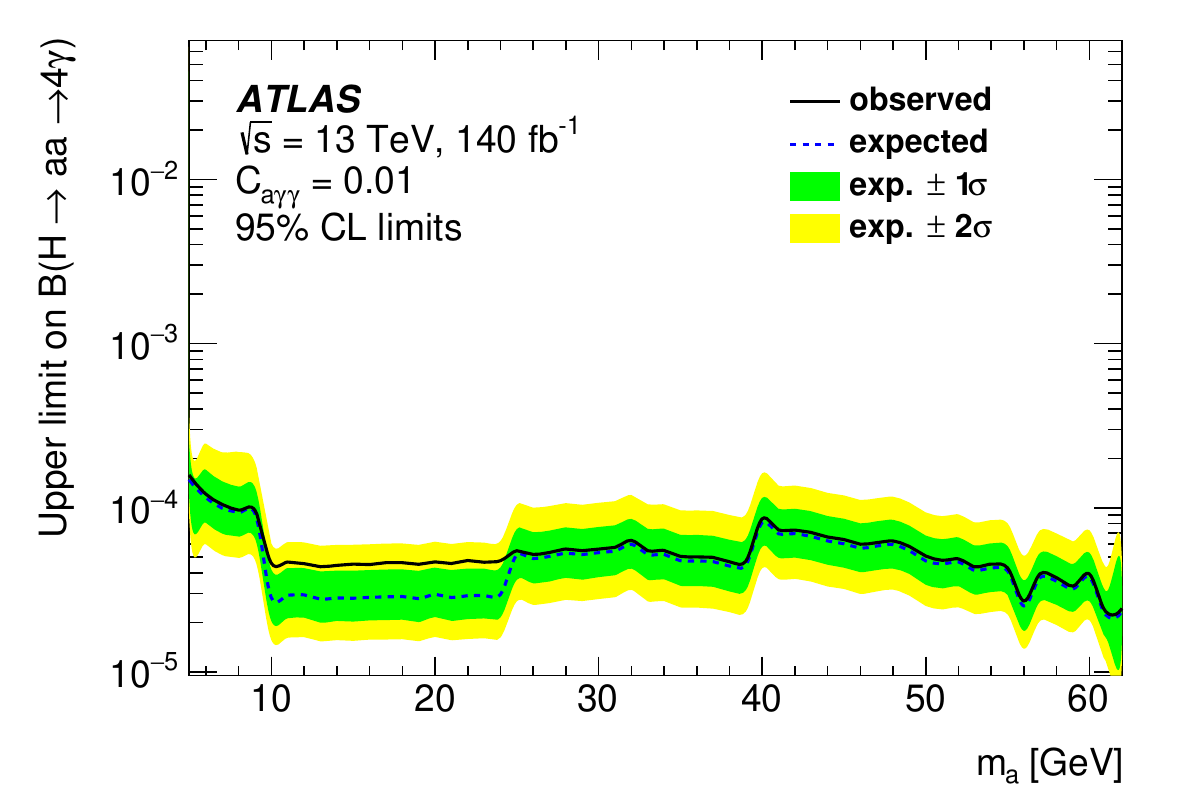}
}
\caption{Zoomed in version of Fig. \ref{fig:1DExclusionLongLived} showing upper
limits on $\mathcal{B}(H\rightarrow aa\rightarrow 4\gamma)$ at 95\% CL as a function of the signal mass hypothesis and for different ALP-photon couplings. (a) $m_a<5.0\,\GeV$, $C_{a\gamma\gamma}=1$; (b) $m_a<5.0\,\GeV$, $C_{a\gamma\gamma}=0.01$; (c)  $m_a > 5.0 \,\GeV$, $C_{a\gamma\gamma}=1$; (d)  $m_a > 5.0 \,\GeV$, $C_{a\gamma\gamma}=0.01$.}
\label{fig:1DExclusionLongLived_split}
\end{figure*}

\begin{figure*}[thb]
\centering
\includegraphics[width=0.69\textwidth]{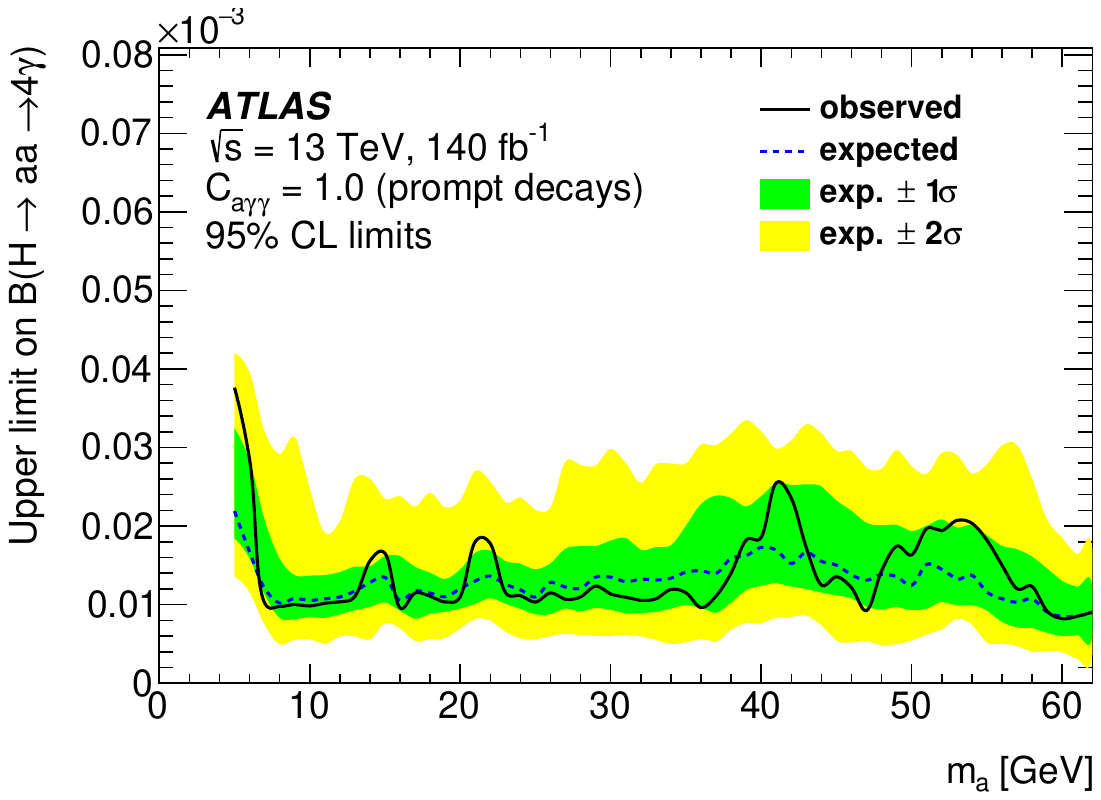}
\caption{
Upper limits on $\mathcal{B}(H\rightarrow aa\rightarrow 4\gamma)$ at 95\% CL as a function of the signal mass hypothesis and for the assumption of promptly decaying ALPs.}
\label{fig:1DExclusionPrompt}
\end{figure*}

The limit on the branching ratio can be converted into a limit on the coupling of axion-like particles to photons, $C_{a\gamma\gamma}$.
The branching ratio, which depends on $C_{a\gamma\gamma}$, can be calculated using the method described in Ref.~\cite{Bauer:2017ris}:
\begin{equation}
\mathcal{B}^{\text{theo}} = \Gamma_{Haa} f_{aa}^2(C_{a\gamma\gamma}) \frac{\mathcal{B}(a\rightarrow\gamma\gamma)}{\Gamma_{H} + \Gamma_{Haa}},
\end{equation}
where it is assumed that all ALPs decay into photons and hence $\mathcal{B}(a\rightarrow\gamma\gamma) = 1$. In the following $m_a$ and $m_H$ refer to the masses of the ALP and Higgs boson, respectively, $v$ is the vacuum expectation value, and the effective coupling of the ALP to the Higgs boson is assumed to be $C_{aH}^{\textrm{eff}}/\Lambda^2 = 1\,\TeV^{-2}$.
$\Gamma_H$ is the total decay width of the Higgs boson. The Higgs boson to ALP decay width $\Gamma_{Haa}$ is calculated as
\begin{equation}
\Gamma_{Haa} = \frac{v^2m_H^3}{32\pi}\frac{\left|C_{aH}^{\textrm{eff}}\right|^2}{\Lambda^4} \left(1-\frac{2m_a^2}{m_H^2}\right)^2 \sqrt{1-\frac{4m_a^2}{m_H^2}}.
\end{equation}
Using the Higgs mass of $m_H = 125\,\GeV$ and assuming $m_a = 10\,\GeV$ and $C_{aH}^{\textrm{eff}}/\Lambda^2 = 1 \, \TeV^{-2}$ yields a branching ratio of the $H\rightarrow aa$ process of $30\%$.
The factor $f_{aa}^2(C_{a\gamma\gamma})$ represents the fraction of ALPs detected inside the detector volume. It depends on the ALP decay length, and hence on $C_{a\gamma\gamma}$, since the decay width is given by
\begin{equation}
\Gamma_{a\gamma\gamma} = \frac{4\pi\alpha^2m_a^3}{\Lambda^2}|C_{a\gamma\gamma}|^2.
\end{equation}
$f_{aa}$ is determined from the signal simulation and interpolated
between the simulated $C_{a\gamma\gamma}$ values.
The coupling is adjusted until the expected branching ratio matches the observed limit on the branching ratio, yielding the corresponding limit on $C_{a\gamma\gamma}$.

The resulting limits are shown in the two-dimensional exclusion plot of $C_{a\gamma\gamma}$ vs. $m_a$ presented in Figure~\ref{fig:2DExclusion}.
This search significantly reduces the allowed parameter space for ALP-based models that could explain the $(g-2)_{\mu}$ discrepancy in the $H\to aa \to 4\gamma$ decay mode, as suggested in Ref.~\cite{Bauer:2017ris}.

\begin{figure*}[thb]
\centering
\includegraphics[width=0.69\textwidth]{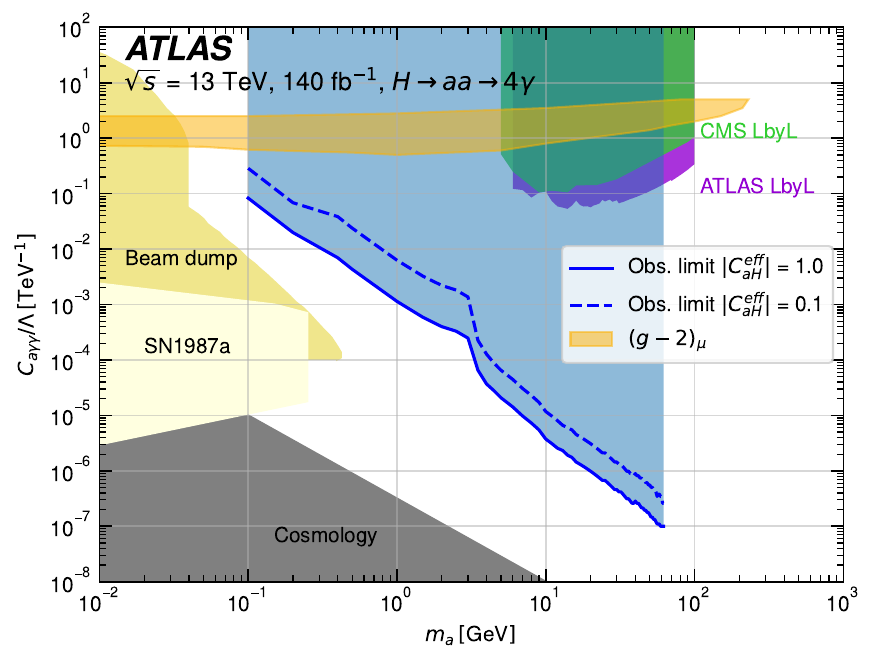}
\caption{
Limits on the ALP mass and coupling to photons at 95\%~CL, assuming $\mathcal{B}(a\rightarrow\gamma\gamma) = 1$,  $\Lambda = 1\,\TeV$ with $|C_{aH}^{\text{eff}}|$ = $1$ (solid line) and $|C_{aH}^{\text{eff}}|$ = $0.1$ (dashed line) as predicted in Ref. \cite{Bauer:2017ris}. The shaded blue area represents the excluded region.
The nearly horizontal orange shaded area indicates the region favoured by an ALP explanation for the $(g-2)_{\mu}$ discrepancy~\cite{Bauer:2017ris}.
Also shown are exclusion limits from the respective ATLAS~\cite{HION-2019-08} and CMS~\cite{CMS-FSQ-16-012} Light-by-Light (LbyL) scattering analysis, and beam dump experiments, supernova SN1987a and cosmological observations adapted from Ref.~\cite{Jaeckel:2015jla}.}
\label{fig:2DExclusion}
\end{figure*}


\FloatBarrier

\section{Conclusion \label{sec:conclusion}}
This paper reports a search for a light pseudoscalar particle ($a$) produced in the decay $H\rightarrow aa$, where $H$ is the 125~\GeV\ Higgs boson. The $a$ boson, which can have a short or long lifetime, decays into two photons, resulting in a final state with four photons with an invariant mass near 125~\GeV.
The analysis uses 140~fb$^{-1}$ of $pp$ collision data at a centre-of-mass energy of 13~\TeV\ collected by the ATLAS detector between 2015 and 2018.
The search aims to identify a narrow $a\rightarrow \gamma\gamma$  resonance with a mass in the range of $100~\MeV$ to $62~\GeV$,
where the resonance decay occurs within a distance of 1970~mm from the collision vertex.
Dedicated search strategies for long-lived $a\to \gamma\gamma$ decays are developed for the first time.
To enable the search for low resonance masses, neural network classifiers are trained to distinguish between single and collimated photon signatures.

No significant excess over the Standard Model backgrounds is observed in the data. The largest deviation from the expected limit, $1.5\sigma$, is observed in the range of $ 10\,\GeV < m_a < 25\,\GeV$. Upper limits at 95\%~CL are set for $\mathcal{B}(H \rightarrow aa \rightarrow 4\gamma)$, which range from $2\times 10^{-5}$ to $3\times 10^{-2}$ depending on $m_a$ for the prompt axion-like particle search. For the search for long-lived ALPs with significant displaced decay vertices, upper limits at 95\%~CL are set, ranging from $2\times 10^{-5}$ to $6\times 10^{-5}$ for $10~\GeV <m_a<62~\GeV$ and from $10^{-4}$ to $3\times 10^{-2}$ for $0.1\,\GeV<m_a<10\,\GeV$. These are the most stringent limits to date.


%
\clearpage
\section*{Acknowledgements}
%

%
%

%
%

We thank CERN for the very successful operation of the LHC and its injectors, as well as the support staff at
CERN and at our institutions worldwide without whom ATLAS could not be operated efficiently.

The crucial computing support from all WLCG partners is acknowledged gratefully, in particular from CERN, the ATLAS Tier-1 facilities at TRIUMF/SFU (Canada), NDGF (Denmark, Norway, Sweden), CC-IN2P3 (France), KIT/GridKA (Germany), INFN-CNAF (Italy), NL-T1 (Netherlands), PIC (Spain), RAL (UK) and BNL (USA), the Tier-2 facilities worldwide and large non-WLCG resource providers. Major contributors of computing resources are listed in Ref.~\cite{ATL-SOFT-PUB-2023-001}.

We gratefully acknowledge the support of ANPCyT, Argentina; YerPhI, Armenia; ARC, Australia; BMWFW and FWF, Austria; ANAS, Azerbaijan; CNPq and FAPESP, Brazil; NSERC, NRC and CFI, Canada; CERN; ANID, Chile; CAS, MOST and NSFC, China; Minciencias, Colombia; MEYS CR, Czech Republic; DNRF and DNSRC, Denmark; IN2P3-CNRS and CEA-DRF/IRFU, France; SRNSFG, Georgia; BMBF, HGF and MPG, Germany; GSRI, Greece; RGC and Hong Kong SAR, China; ISF and Benoziyo Center, Israel; INFN, Italy; MEXT and JSPS, Japan; CNRST, Morocco; NWO, Netherlands; RCN, Norway; MEiN, Poland; FCT, Portugal; MNE/IFA, Romania; MESTD, Serbia; MSSR, Slovakia; ARRS and MIZ\v{S}, Slovenia; DSI/NRF, South Africa; MICINN, Spain; SRC and Wallenberg Foundation, Sweden; SERI, SNSF and Cantons of Bern and Geneva, Switzerland; MOST, Taipei; TENMAK, T\"urkiye; STFC, United Kingdom; DOE and NSF, United States of America.

Individual groups and members have received support from BCKDF, CANARIE, CRC and DRAC, Canada; CERN-CZ, PRIMUS 21/SCI/017 and UNCE SCI/013, Czech Republic; COST, ERC, ERDF, Horizon 2020, ICSC-NextGenerationEU and Marie Sk{\l}odowska-Curie Actions, European Union; Investissements d'Avenir Labex, Investissements d'Avenir Idex and ANR, France; DFG and AvH Foundation, Germany; Herakleitos, Thales and Aristeia programmes co-financed by EU-ESF and the Greek NSRF, Greece; BSF-NSF and MINERVA, Israel; Norwegian Financial Mechanism 2014-2021, Norway; NCN and NAWA, Poland; La Caixa Banking Foundation, CERCA Programme Generalitat de Catalunya and PROMETEO and GenT Programmes Generalitat Valenciana, Spain; G\"{o}ran Gustafssons Stiftelse, Sweden; The Royal Society and Leverhulme Trust, United Kingdom.

In addition, individual members wish to acknowledge support from Chile: Agencia Nacional de Investigaci\'on y Desarrollo (FONDECYT 1190886, FONDECYT 1210400, FONDECYT 1230812, FONDECYT 1230987); China: National Natural Science Foundation of China (NSFC - 12175119, NSFC 12275265, NSFC-12075060); Czech Republic: PRIMUS Research Programme (PRIMUS/21/SCI/017); EU: H2020 European Research Council (ERC - 101002463); European Union: European Research Council (ERC - 948254), Horizon 2020 Framework Programme (MUCCA - CHIST-ERA-19-XAI-00), European Union, Future Artificial Intelligence Research (FAIR-NextGenerationEU PE00000013), Italian Center for High Performance Computing, Big Data and Quantum Computing (ICSC, NextGenerationEU), Marie Sklodowska-Curie Actions (EU H2020 MSC IF GRANT NO 101033496); France: Agence Nationale de la Recherche (ANR-20-CE31-0013, ANR-21-CE31-0013, ANR-21-CE31-0022), Investissements d'Avenir Idex (ANR-11-LABX-0012), Investissements d'Avenir Labex (ANR-11-LABX-0012); Germany: Baden-Württemberg Stiftung (BW Stiftung-Postdoc Eliteprogramme), Deutsche Forschungsgemeinschaft (DFG - CR 312/5-1); Italy: Istituto Nazionale di Fisica Nucleare (FELLINI G.A. n. 754496, ICSC, NextGenerationEU); Japan: Japan Society for the Promotion of Science (JSPS KAKENHI JP21H05085, JSPS KAKENHI JP22H01227, JSPS KAKENHI JP22H04944); Netherlands: Netherlands Organisation for Scientific Research (NWO Veni 2020 - VI.Veni.202.179); Norway: Research Council of Norway (RCN-314472); Poland: Polish National Agency for Academic Exchange (PPN/PPO/2020/1/00002/U/00001), Polish National Science Centre (NCN 2021/42/E/ST2/00350, NCN UMO-2019/34/E/ST2/00393, UMO-2020/37/B/ST2/01043, UMO-2021/40/C/ST2/00187); Slovenia: Slovenian Research Agency (ARIS grant J1-3010); Spain: BBVA Foundation (LEO22-1-603), Generalitat Valenciana (Artemisa, FEDER, IDIFEDER/2018/048), La Caixa Banking Foundation (LCF/BQ/PI20/11760025), Ministry of Science and Innovation (MCIN \& NextGenEU PCI2022-135018-2, MICIN \& FEDER PID2021-125273NB, RYC2019-028510-I, RYC2020-030254-I, RYC2021-031273-I, RYC2022-038164-I), PROMETEO and GenT Programmes Generalitat Valenciana (CIDEGENT/2019/023, CIDEGENT/2019/027); Sweden: Swedish Research Council (VR 2018-00482, VR 2022-03845, VR 2022-04683, VR grant 2021-03651), Knut and Alice Wallenberg Foundation (KAW 2017.0100, KAW 2018.0157, KAW 2018.0458, KAW 2019.0447); Switzerland: Swiss National Science Foundation (SNSF - PCEFP2\_194658); United Kingdom: Leverhulme Trust (Leverhulme Trust RPG-2020-004); United States of America: U.S. Department of Energy (ECA DE-AC02-76SF00515), Neubauer Family Foundation.

%
%


%
%
%
%
%
%

%
%

%
%
%
\printbibliography

@article{Grzadkowski:2018ohf,
    author = "Grzadkowski, Bohdan and Haber, Howard E. and Ogreid, Odd Magne and Osland, Per",
    title = "{Heavy Higgs boson decays in the alignment limit of the 2HDM}",
    eprint = "1808.01472",
    archivePrefix = "arXiv",
    primaryClass = "hep-ph",
    reportNumber = "SCIPP-18/05, SCIPP-18-05",
    doi = "10.1007/JHEP12(2018)056",
    journal = "JHEP",
    volume = "12",
    pages = "056",
    year = "2018"
}

@software{pyhf,
  author = {Lukas Heinrich and Matthew Feickert and Giordon Stark},
  title = "{pyhf: v0.7.1}",
  version = {0.7.1},
  doi = {10.5281/zenodo.1169739},
  url = {https://doi.org/10.5281/zenodo.1169739},
  note = {https://github.com/scikit-hep/pyhf/releases/tag/v0.7.1}
}

@article{Muong-2:2023cdq,
    author = "Aguillard, D. P. and others",
    collaboration = "Muon g-2",
    title = "{Measurement of the Positive Muon Anomalous Magnetic Moment to 0.20 ppm}",
    eprint = "2308.06230",
    archivePrefix = "arXiv",
    primaryClass = "hep-ex",
    reportNumber = "FERMILAB-PUB-23-385-AD-CSAID-PPD",
    month = "8",
    year = "2023"
}

@article{ATLAS:2023mnw,
    author         = "{ATLAS Collaboration}",
    collaboration = "ATLAS",
    title = "{Electron and photon energy calibration with the ATLAS detector using LHC Run 2 data}",
    eprint = "2309.05471",
    archivePrefix = "arXiv",
    primaryClass = "hep-ex",
    reportNumber = "CERN-EP-2023-128",
    month = "9",
    year = "2023"
}

@article{ATLAS:2023dxj,
    author         = "{ATLAS Collaboration}",
    collaboration = "ATLAS",
    title = "{Electron and photon efficiencies in LHC Run 2 with the ATLAS experiment}",
    eprint = "2308.13362",
    archivePrefix = "arXiv",
    primaryClass = "hep-ex",
    reportNumber = "CERN-EP-2023-182",
    month = "8",
    year = "2023"
}

@article{Jaeckel:2015jla,
    author = "Jaeckel, Joerg and Spannowsky, Michael",
    title = "{Probing MeV to 90 GeV axion-like particles with LEP and LHC}",
    eprint = "1509.00476",
    archivePrefix = "arXiv",
    primaryClass = "hep-ph",
    doi = "10.1016/j.physletb.2015.12.037",
    journal = "Phys. Lett. B",
    volume = "753",
    pages = "482--487",
    year = "2016"
}

@article{ATLAS:2023tkt,
    collaboration = "ATLAS",
    author         = "{ATLAS Collaboration}",
title = "{Combination of searches for invisible decays of the Higgs boson using 139 fb$^{-1}$ of proton-proton collision data at $\sqrt{s}=13$ TeV collected with the ATLAS experiment}",
    eprint = "2301.10731",
    archivePrefix = "arXiv",
    primaryClass = "hep-ex",
    reportNumber = "CERN-EP-2022-289",
    doi = "10.1016/j.physletb.2023.137963",
    journal = "Phys. Lett. B",
    volume = "842",
    pages = "137963",
    year = "2023"
}

@article{ATLAS:2022hro,
    collaboration = "ATLAS",
    author         = "{ATLAS Collaboration}",
    title = "{Luminosity determination in $pp$ collisions at $\sqrt{s}=13$ TeV using the ATLAS detector at the LHC}",
    eprint = "2212.09379",
    archivePrefix = "arXiv",
    primaryClass = "hep-ex",
    reportNumber = "CERN-EP-2022-281",
    doi = "10.1140/epjc/s10052-023-11747-w",
    journal = "Eur. Phys. J. C",
    volume = "83",
    number = "10",
    pages = "982",
    year = "2023"
}

@article{CMS:2022dwd,
    author = "{CMS Collaboration}",
    collaboration = "CMS",
    title = "{A portrait of the Higgs boson by the CMS experiment ten years after the discovery.}",
    eprint = "2207.00043",
    archivePrefix = "arXiv",
    primaryClass = "hep-ex",
    reportNumber = "CMS-HIG-22-001, CERN-EP-2022-039",
    doi = "10.1038/s41586-022-04892-x",
    journal = "Nature",
    volume = "607",
    number = "7917",
    pages = "60--68",
    year = "2022"
}

@article{ATLAS:2022vkf,
    author = "{ATLAS Collaboration}",
    collaboration = "ATLAS",
    title = "{A detailed map of Higgs boson interactions by the ATLAS experiment ten years after the discovery}",
    eprint = "2207.00092",
    archivePrefix = "arXiv",
    primaryClass = "hep-ex",
    reportNumber = "CERN-EP-2022-057",
    doi = "10.1038/s41586-022-04893-w",
    journal = "Nature",
    volume = "607",
    number = "7917",
    pages = "52--59",
    year = "2022",
    related = "ATLAS:2022vkf-err",
    relatedstring="Erratum:"
}

@article{Avoni:2018iuv,
    author = "Avoni, G. and others",
    title = "{The new LUCID-2 detector for luminosity measurement and monitoring in ATLAS}",
    doi = "10.1088/1748-0221/13/07/P07017",
    journal = "JINST",
    volume = "13",
    number = "07",
    pages = "P07017",
    year = "2018"
}

@article{Agostinelli:2002hh,
    author = "Agostinelli, S. and others",
    collaboration = "GEANT4",
    title = "{GEANT4--a simulation toolkit}",
    reportNumber = "SLAC-PUB-9350, FERMILAB-PUB-03-339, CERN-IT-2002-003",
    doi = "10.1016/S0168-9002(03)01368-8",
    journal = "Nucl. Instrum. Meth. A",
    volume = "506",
    pages = "250--303",
    year = "2003"
}

@article{Frixione:2007vw,
    author = "Frixione, Stefano and Nason, Paolo and Oleari, Carlo",
    title = "{Matching NLO QCD computations with Parton Shower simulations: the POWHEG method}",
    eprint = "0709.2092",
    archivePrefix = "arXiv",
    primaryClass = "hep-ph",
    reportNumber = "BICOCCA-FT-07-9, GEF-TH-21-2007",
    doi = "10.1088/1126-6708/2007/11/070",
    journal = "JHEP",
    volume = "11",
    pages = "070",
    year = "2007"
}

@article{Alioli:2010xd,
    author = "Alioli, Simone and Nason, Paolo and Oleari, Carlo and Re, Emanuele",
    title = "{A general framework for implementing NLO calculations in shower Monte Carlo programs: the POWHEG BOX}",
    eprint = "1002.2581",
    archivePrefix = "arXiv",
    primaryClass = "hep-ph",
    reportNumber = "DESY-10-018, SFB-CPP-10-22, IPPP-10-11, DCPT-10-22",
    doi = "10.1007/JHEP06(2010)043",
    journal = "JHEP",
    volume = "06",
    pages = "043",
    year = "2010"
}

@article{Sjostrand:2014zea,
    author = {Sj\"ostrand, Torbj\"orn and Ask, Stefan and Christiansen, Jesper R. and Corke, Richard and Desai, Nishita and Ilten, Philip and Mrenna, Stephen and Prestel, Stefan and Rasmussen, Christine O. and Skands, Peter Z.},
    title = "{An introduction to PYTHIA 8.2}",
    eprint = "1410.3012",
    archivePrefix = "arXiv",
    primaryClass = "hep-ph",
    reportNumber = "LU-TP-14-36, MCNET-14-22, CERN-PH-TH-2014-190, FERMILAB-PUB-14-316-CD, DESY-14-178, SLAC-PUB-16122",
    doi = "10.1016/j.cpc.2015.01.024",
    journal = "Comput. Phys. Commun.",
    volume = "191",
    pages = "159--177",
    year = "2015"
}

@article{deFlorian:2016spz,
    author = "de Florian, D. and others",
    collaboration = "LHC Higgs Cross Section Working Group",
    title = "{Handbook of LHC Higgs Cross Sections: 4. Deciphering the Nature of the Higgs Sector}",
    eprint = "1610.07922",
    archivePrefix = "arXiv",
    primaryClass = "hep-ph",
    reportNumber = "CERN-2017-002-M, CERN-2017-002",
    doi = "10.23731/CYRM-2017-002",
    volume = "2/2017",
    month = "10",
    year = "2016"
}

@article{Anastasiou:2016cez,
    author = "Anastasiou, Charalampos and Duhr, Claude and Dulat, Falko and Furlan, Elisabetta and Gehrmann, Thomas and Herzog, Franz and Lazopoulos, Achilleas and Mistlberger, Bernhard",
    title = "{High precision determination of the gluon fusion Higgs boson cross-section at the LHC}",
    eprint = "1602.00695",
    archivePrefix = "arXiv",
    primaryClass = "hep-ph",
    reportNumber = "CP3-16-01, ZU-TH-27-15, NIKHEF-2016-004, CERN-TH-2016-006",
    doi = "10.1007/JHEP05(2016)058",
    journal = "JHEP",
    volume = "05",
    pages = "058",
    year = "2016"
}

@article{Actis:2008ug,
    author = "Actis, Stefano and Passarino, Giampiero and Sturm, Christian and Uccirati, Sandro",
    title = "{NLO Electroweak Corrections to Higgs Boson Production at Hadron Colliders}",
    eprint = "0809.1301",
    archivePrefix = "arXiv",
    primaryClass = "hep-ph",
    reportNumber = "PITHA-08-20, SFB-CPP-08-62, TTP08-38",
    doi = "10.1016/j.physletb.2008.10.018",
    journal = "Phys. Lett. B",
    volume = "670",
    pages = "12--17",
    year = "2008"
}

@article{Anastasiou:2008tj,
    author = "Anastasiou, Charalampos and Boughezal, Radja and Petriello, Frank",
    title = "{Mixed QCD-electroweak corrections to Higgs boson production in gluon fusion}",
    eprint = "0811.3458",
    archivePrefix = "arXiv",
    primaryClass = "hep-ph",
    doi = "10.1088/1126-6708/2009/04/003",
    journal = "JHEP",
    volume = "04",
    pages = "003",
    year = "2009"
}

@article{ATLAS:2017lpx,
    author = "Aaboud, Morad and others",
    collaboration = "ATLAS",
    title = "{Measurement of the production cross section of three isolated photons in $pp$ collisions at $\sqrt{s}$ = 8 TeV using the ATLAS detector}",
    eprint = "1712.07291",
    archivePrefix = "arXiv",
    primaryClass = "hep-ex",
    reportNumber = "CERN-EP-2017-302",
    doi = "10.1016/j.physletb.2018.03.057",
    journal = "Phys. Lett. B",
    volume = "781",
    pages = "55--76",
    year = "2018"
}

@article{Bauer:2017ris,
    author = "Bauer, Martin and Neubert, Matthias and Thamm, Andrea",
    title = "{Collider Probes of Axion-Like Particles}",
    eprint = "1708.00443",
    archivePrefix = "arXiv",
    primaryClass = "hep-ph",
    reportNumber = "MITP-17-047",
    doi = "10.1007/JHEP12(2017)044",
    journal = "JHEP",
    volume = "12",
    pages = "044",
    year = "2017"
}

@article{Muong-2:2021ojo,
    author = "{Muon g-2 Collaboration}",
    collaboration = "Muon g-2",
    title = "{Measurement of the Positive Muon Anomalous Magnetic Moment to 0.46 ppm}",
    eprint = "2104.03281",
    archivePrefix = "arXiv",
    primaryClass = "hep-ex",
    reportNumber = "FERMILAB-PUB-21-132-E",
    doi = "10.1103/PhysRevLett.126.141801",
    journal = "Phys. Rev. Lett.",
    volume = "126",
    number = "14",
    pages = "141801",
    year = "2021"
}

@article{Muong-2:2006rrc,
    author = "{Muon g-2 Collaboration}",
    collaboration = "Muon g-2",
    title = "{Final report of the E821 muon anomalous magnetic moment measurement at BNL}",
    eprint = "hep-ex/0602035",
    archivePrefix = "arXiv",
    doi = "10.1103/PhysRevD.73.072003",
    journal = "Phys. Rev. D",
    volume = "73",
    pages = "072003",
    year = "2006"
}

@article{Aoyama:2020ynm,
    author = "Aoyama, T. and others",
    title = "{The anomalous magnetic moment of the muon in the Standard Model}",
    eprint = "2006.04822",
    archivePrefix = "arXiv",
    primaryClass = "hep-ph",
    reportNumber = "FERMILAB-PUB-20-207-T, INT-PUB-20-021, KEK Preprint 2020-5,
  MITP/20-028, KEK Preprint 2020-5, MITP/20-028, CERN-TH-2020-075, IFT-UAM/CSIC-20-74, LMU-ASC 18/20, LTH 1234,
  LU TP 20-20, LTH 1234, LU TP 20-20, MAN/HEP/2020/003, PSI-PR-20-06, UWThPh 2020-14, ZU-TH 18/20",
    doi = "10.1016/j.physrep.2020.07.006",
    journal = "Phys. Rept.",
    volume = "887",
    pages = "1--166",
    year = "2020"
}

@article{RBC:2018dos,
    author = "{RBC and UKQCD Collaborations}",
    collaboration = "RBC, UKQCD",
    title = "{Calculation of the hadronic vacuum polarization contribution to the muon anomalous magnetic moment}",
    eprint = "1801.07224",
    archivePrefix = "arXiv",
    primaryClass = "hep-lat",
    doi = "10.1103/PhysRevLett.121.022003",
    journal = "Phys. Rev. Lett.",
    volume = "121",
    number = "2",
    pages = "022003",
    year = "2018"
}

@article{Dobrescu:2000yn,
    author = "Dobrescu, Bogdan A. and Matchev, Konstantin T.",
    title = "{Light axion within the next-to-minimal supersymmetric standard model}",
    eprint = "hep-ph/0008192",
    archivePrefix = "arXiv",
    reportNumber = "FERMILAB-PUB-00-134-T",
    doi = "10.1088/1126-6708/2000/09/031",
    journal = "JHEP",
    volume = "09",
    pages = "031",
    year = "2000"
}

@article{Ellwanger:2003jt,
    author = "Ellwanger, Ulrich and Gunion, John F. and Hugonie, Cyril and Moretti, Stefano",
    title = "{Towards a No-Lose Theorem for NMSSM Higgs Discovery at the LHC}",
    eprint = "hep-ph/0305109",
    archivePrefix = "arXiv",
    reportNumber = "UCD-03-4, IFIC-03-19, LPT-03-35, SHEP-03-08",
    month = "5",
    year = "2003"
}

@article{Profumo:2007wc,
    author = "Profumo, Stefano and Ramsey-Musolf, Michael J. and Shaughnessy, Gabe",
    title = "{Singlet Higgs phenomenology and the electroweak phase transition}",
    eprint = "0705.2425",
    archivePrefix = "arXiv",
    primaryClass = "hep-ph",
    reportNumber = "CALTECH-MAP-333, MADPH-07-1489",
    doi = "10.1088/1126-6708/2007/08/010",
    journal = "JHEP",
    volume = "08",
    pages = "010",
    year = "2007"
}

@article{Blinov:2015sna,
    author = "Blinov, Nikita and Kozaczuk, Jonathan and Morrissey, David E. and Tamarit, Carlos",
    title = "{Electroweak Baryogenesis from Exotic Electroweak Symmetry Breaking}",
    eprint = "1504.05195",
    archivePrefix = "arXiv",
    primaryClass = "hep-ph",
    reportNumber = "IPPP-15-23, DCPT-15-46",
    doi = "10.1103/PhysRevD.92.035012",
    journal = "Phys. Rev. D",
    volume = "92",
    number = "3",
    pages = "035012",
    year = "2015"
}

@article{Craig:2015pha,
    author = "Craig, Nathaniel and Katz, Andrey and Strassler, Matt and Sundrum, Raman",
    title = "{Naturalness in the Dark at the LHC}",
    eprint = "1501.05310",
    archivePrefix = "arXiv",
    primaryClass = "hep-ph",
    reportNumber = "UMD-PP-014-028, CERN-PH-TH-2014-263",
    doi = "10.1007/JHEP07(2015)105",
    journal = "JHEP",
    volume = "07",
    pages = "105",
    year = "2015"
}

@article{Curtin:2015fna,
    author = "Curtin, David and Verhaaren, Christopher B.",
    title = "{Discovering Uncolored Naturalness in Exotic Higgs Decays}",
    eprint = "1506.06141",
    archivePrefix = "arXiv",
    primaryClass = "hep-ph",
    doi = "10.1007/JHEP12(2015)072",
    journal = "JHEP",
    volume = "12",
    pages = "072",
    year = "2015"
}

@article{Silveira:1985rk,
    author = "Silveira, Vanda and Zee, A.",
    title = "{SCALAR PHANTOMS}",
    reportNumber = "DOE-ER-40048-13 P5",
    doi = "10.1016/0370-2693(85)90624-0",
    journal = "Phys. Lett. B",
    volume = "161",
    pages = "136--140",
    year = "1985"
}

@article{Pospelov:2007mp,
    author = "Pospelov, Maxim and Ritz, Adam and Voloshin, Mikhail",
    title = "{Secluded WIMP Dark Matter}",
    eprint = "0711.4866",
    archivePrefix = "arXiv",
    primaryClass = "hep-ph",
    doi = "10.1016/j.physletb.2008.02.052",
    journal = "Phys. Lett. B",
    volume = "662",
    pages = "53--61",
    year = "2008"
}

@article{Draper:2010ew,
    author = "Draper, Patrick and Liu, Tao and Wagner, Carlos E. M. and Wang, Lian-Tao and Zhang, Hao",
    title = "{Dark Light-Higgs Bosons}",
    eprint = "1009.3963",
    archivePrefix = "arXiv",
    primaryClass = "hep-ph",
    reportNumber = "ANL-HEP-PR-10-48, EFI-10-22",
    doi = "10.1103/PhysRevLett.106.121805",
    journal = "Phys. Rev. Lett.",
    volume = "106",
    pages = "121805",
    year = "2011"
}

@article{Ipek:2014gua,
    author = "Ipek, Seyda and McKeen, David and Nelson, Ann E.",
    title = "{Renormalizable model for the Galactic Center gamma-ray excess from dark matter annihilation}",
    eprint = "1404.3716",
    archivePrefix = "arXiv",
    primaryClass = "hep-ph",
    doi = "10.1103/PhysRevD.90.055021",
    journal = "Phys. Rev. D",
    volume = "90",
    number = "5",
    pages = "055021",
    year = "2014"
}

@article{Martin:2014sxa,
    author = "Martin, Adam and Shelton, Jessie and Unwin, James",
    title = "{Fitting the Galactic Center Gamma-Ray Excess with Cascade Annihilations}",
    eprint = "1405.0272",
    archivePrefix = "arXiv",
    primaryClass = "hep-ph",
    doi = "10.1103/PhysRevD.90.103513",
    journal = "Phys. Rev. D",
    volume = "90",
    number = "10",
    pages = "103513",
    year = "2014"
}

@article{Boehm:2014hva,
    author = "Boehm, C\'eline and Dolan, Matthew J. and McCabe, Christopher and Spannowsky, Michael and Wallace, Chris J.",
    title = "{Extended gamma-ray emission from Coy Dark Matter}",
    eprint = "1401.6458",
    archivePrefix = "arXiv",
    primaryClass = "hep-ph",
    reportNumber = "IPPP-13-31, DCPT-13-62, SLAC-PUB-15893",
    doi = "10.1088/1475-7516/2014/05/009",
    journal = "JCAP",
    volume = "05",
    pages = "009",
    year = "2014"
}

@article{Curtin:2013fra,
    author = "Curtin, David and others",
    title = "{Exotic decays of the 125 GeV Higgs boson}",
    eprint = "1312.4992",
    archivePrefix = "arXiv",
    primaryClass = "hep-ph",
    reportNumber = "YITP-13-47, PITT-PACC-1314",
    doi = "10.1103/PhysRevD.90.075004",
    journal = "Phys. Rev. D",
    volume = "90",
    number = "7",
    pages = "075004",
    year = "2014"
}

@article{Evans:2008zzb,
    editor = "Evans, Lyndon and Bryant, Philip",
    title = "{LHC Machine}",
    doi = "10.1088/1748-0221/3/08/S08001",
    journal = "JINST",
    volume = "3",
    pages = "S08001",
    year = "2008"
}

@article{Heinrich:2021gyp,
    author = "Heinrich, Lukas and Feickert, Matthew and Stark, Giordon and Cranmer, Kyle",
    title = "{pyhf: pure-Python implementation of HistFactory statistical models}",
    doi = "10.21105/joss.02823",
    journal = "J. Open Source Softw.",
    volume = "6",
    number = "58",
    pages = "2823",
    year = "2021"
}

@article{Read:2002hq,
    author = "Read, Alexander L.",
    editor = "Whalley, M. R. and Lyons, L.",
    title = "{Presentation of search results: The CL(s) technique}",
    doi = "10.1088/0954-3899/28/10/313",
    journal = "J. Phys. G",
    volume = "28",
    pages = "2693--2704",
    year = "2002"
}

@article{CMS:2022xxa,
    collaboration = "CMS",
    author         = "{CMS Collaboration}",
    title = "{Search for the exotic decay of the Higgs boson into two light pseudoscalars with four photons in the final state in proton-proton collisions at $\sqrt{s}$ = 13 TeV}",
    eprint = "2208.01469",
    archivePrefix = "arXiv",
    primaryClass = "hep-ex",
    reportNumber = "CMS-HIG-21-003, CERN-EP-2022-095",
    doi = "10.1007/JHEP07(2023)148",
    journal = "JHEP",
    volume = "07",
    pages = "148",
    year = "2023"
}

@article{CMS:2022fyt,
    collaboration = "CMS",
    author         = "{CMS Collaboration}",
    title = "{Search for exotic Higgs boson decays H $\to$$\mathcal{A}\mathcal{A}$$\to$ 4$\gamma$ with events containing two merged diphotons in proton-proton collisions at $\sqrt{s}$ = 13 TeV}",
    eprint = "2209.06197",
    archivePrefix = "arXiv",
    primaryClass = "hep-ex",
    reportNumber = "CMS-HIG-21-016, CERN-EP-2022-151",
    doi = "10.1103/PhysRevLett.131.101801",
    journal = "Phys. Rev. Lett.",
    volume = "131",
    pages = "101801",
    year = "2023"
}

@article{ATLAS:2015rsn,
    author = "{ATLAS Collaboration}",
    collaboration = "ATLAS",
    title = "{Search for new phenomena in events with at least three photons collected in $pp$ collisions at $\sqrt{s}$ = 8 TeV with the ATLAS detector}",
    eprint = "1509.05051",
    archivePrefix = "arXiv",
    primaryClass = "hep-ex",
    reportNumber = "CERN-PH-EP-2015-187",
    doi = "10.1140/epjc/s10052-016-4034-8",
    journal = "Eur. Phys. J. C",
    volume = "76",
    number = "4",
    pages = "210",
    year = "2016"
}

@article{Baglio:2020wgt,
    author = {Baglio, J. and Campanario, F. and Glaus, S. and M\"uhlleitner, M. and Ronca, J. and Spira, M.},
    title = "{$gg\to HH$ : Combined uncertainties}",
    eprint = "2008.11626",
    archivePrefix = "arXiv",
    primaryClass = "hep-ph",
    reportNumber = "CERN-TH-2020-139, IFIC/20-42, FTUV-20-0823, KA-TP-11-2020,
  PSI-PR-20-13, KA-TP-11-2020, PSI-PR-20-13",
    doi = "10.1103/PhysRevD.103.056002",
    journal = "Phys. Rev. D",
    volume = "103",
    number = "5",
    pages = "056002",
    year = "2021"
}

@article{Stump:2003yu,
    author = "Stump, Daniel and Huston, Joey and Pumplin, Jon and Tung, Wu-Ki and Lai, H. L. and Kuhlmann, Steve and Owens, J. F.",
    title = "{Inclusive jet production, parton distributions, and the search for new physics}",
    eprint = "hep-ph/0303013",
    archivePrefix = "arXiv",
    reportNumber = "MSU-HEP-030303",
    doi = "10.1088/1126-6708/2003/10/046",
    journal = "JHEP",
    volume = "10",
    pages = "046",
    year = "2003"
}

@article{Ball:2012cx,
    author = "{NNPDF Collaboration}",
    title = "{Parton distributions with LHC data}",
    eprint = "1207.1303",
    archivePrefix = "arXiv",
    primaryClass = "hep-ph",
    reportNumber = "EDINBURGH-2012-08, IFUM-FT-997, FR-PHENO-2012-014, RWTH-TTK-12-25, CERN-PH-TH-2012-037, SFB-CPP-12-47",
    doi = "10.1016/j.nuclphysb.2012.10.003",
    journal = "Nucl. Phys. B",
    volume = "867",
    pages = "244--289",
    year = "2013"
}

@article{Ball:2014uwa,
      author         = "{NNPDF Collaboration} and Ball, Richard D. and others",
      title          = "{Parton distributions for the LHC run II}",
      journal        = "JHEP",
      volume         = "04",
      year           = "2015",
      pages          = "040",
      doi            = "10.1007/JHEP04(2015)040",
      eprint         = "1410.8849",
      archivePrefix  = "arXiv",
      primaryClass   = "hep-ph",
      reportNumber   = "EDINBURGH-2014-15, IFUM-1034-FT, CERN-PH-TH-2013-253,
                        OUTP-14-11P, CAVENDISH-HEP-14-11",
      SLACcitation   = "%%CITATION = ARXIV:1410.8849;%%"
}

@Article{Nason:2004rx,
      author         = "Nason, Paolo",
      title          = "{A new method for combining NLO QCD with shower Monte Carlo algorithms}",
      journal        = "JHEP",
      volume         = "11",
      pages          = "040",
      doi            = "10.1088/1126-6708/2004/11/040",
      year           = "2004",
      eprint         = "hep-ph/0409146",
      archivePrefix  = "arXiv",
}

@Article{Gleisberg:2008ta,
      author         = "Gleisberg, T. and H{\"o}che, S. and Krauss, F. and Sch{\"o}nherr, M. and Schumann, S. and Siegert, F. and Winter J.",
      title          = "{Event generation with SHERPA 1.1}",
      journal        = "JHEP",
      volume         = "02",
      pages          = "007",
      doi            = "10.1088/1126-6708/2009/02/007",
      year           = "2009",
      eprint         = "0811.4622",
      archivePrefix  = "arXiv",
      primaryClass   = "hep-ph",
}

@Article{Hoeche:2009rj,
      author         = "H{\"o}che, Stefan and Krauss, Frank and Schumann, Steffen and Siegert, Frank",
      title          = "{QCD matrix elements and truncated showers}",
      journal        = "JHEP",
      volume         = "05",
      pages          = "053",
      doi            = "10.1088/1126-6708/2009/05/053",
      year           = "2009",
      eprint         = "0903.1219",
      archivePrefix  = "arXiv",
      primaryClass   = "hep-ph",
}

@article{Bothmann:2019yzt,
      author = "Bothmann, Enrico and others",
      title = "{Event generation with Sherpa 2.2}",
      journal = "SciPost Phys.",
      volume = "7",
      year = "2019",
      number = "3",
      pages = "034",
      doi = "10.21468/SciPostPhys.7.3.034",
      reportNumber = "FERMILAB-PUB-19-218-T, SLAC-PUB-17433, IPPP/19/42, MCNET-19-11",
      eprint = "1905.09127",
      archivePrefix = "arXiv",
      primaryClass = "hep-ph",
}

@Article{PERF-2007-01,
    author         = "{ATLAS Collaboration}",
    title          = "{The ATLAS Experiment at the CERN Large Hadron Collider}",
    journal        = "JINST",
    volume         = "3",
    year           = "2008",
    pages          = "S08003",
    doi            = "10.1088/1748-0221/3/08/S08003",
    primaryClass   = "hep-ex",
}

@Article{SOFT-2010-01,
    author         = "{ATLAS Collaboration}",
    title          = "{The ATLAS Simulation Infrastructure}",
    journal        = "Eur. Phys. J. C",
    volume         = "70",
    year           = "2010",
    pages          = "823",
    doi            = "10.1140/epjc/s10052-010-1429-9",
    eprint         = "1005.4568",
    archivePrefix  = "arXiv",
    primaryClass   = "physics.ins-det",
}

@Article{HIGG-2012-27,
    author         = "{ATLAS Collaboration}",
    title          = "{Observation of a new particle in the search for the Standard Model Higgs boson with the ATLAS detector at the LHC}",
    journal        = "Phys. Lett. B",
    volume         = "716",
    year           = "2012",
    pages          = "1",
    doi            = "10.1016/j.physletb.2012.08.020",
    reportNumber   = "CERN-PH-EP-2012-218",
    eprint         = "1207.7214",
    archivePrefix  = "arXiv",
    primaryClass   = "hep-ex",
}

@Article{STDM-2012-23,
    author         = "{ATLAS Collaboration}",
    title          = "{Measurement of the \(Z/\gamma^*\) boson transverse momentum distribution in \(pp\) collisions at \(\sqrt{s} = 7\,\text{TeV}\) with the ATLAS detector}",
    journal        = "JHEP",
    volume         = "09",
    year           = "2014",
    pages          = "145",
    doi            = "10.1007/JHEP09(2014)145",
    reportNumber   = "CERN-PH-EP-2014-075",
    eprint         = "1406.3660",
    archivePrefix  = "arXiv",
    primaryClass   = "hep-ex",
}

@Article{TRIG-2016-01,
    author         = "{ATLAS Collaboration}",
    title          = "{Performance of the ATLAS trigger system in 2015}",
    journal        = "Eur. Phys. J. C",
    volume         = "77",
    year           = "2017",
    pages          = "317",
    doi            = "10.1140/epjc/s10052-017-4852-3",
    reportNumber   = "CERN-EP-2016-241",
    eprint         = "1611.09661",
    archivePrefix  = "arXiv",
    primaryClass   = "hep-ex",
}

@Article{DAPR-2018-01,
    author         = "{ATLAS Collaboration}",
    title          = "{ATLAS data quality operations and performance for 2015--2018 data-taking}",
    journal        = "JINST",
    volume         = "15",
    year           = "2020",
    pages          = "P04003",
    doi            = "10.1088/1748-0221/15/04/P04003",
    reportNumber   = "CERN-EP-2019-207",
    eprint         = "1911.04632",
    archivePrefix  = "arXiv",
    primaryClass   = "physics.ins-det",
}

@Article{EGAM-2018-01,
    author         = "{ATLAS Collaboration}",
    title          = "{Electron and photon performance measurements with the ATLAS detector using the 2015--2017 LHC proton--proton collision data}",
    journal        = "JINST",
    volume         = "14",
    year           = "2019",
    pages          = "P12006",
    doi            = "10.1088/1748-0221/14/12/P12006",
    reportNumber   = "CERN-EP-2019-145",
    eprint         = "1908.00005",
    archivePrefix  = "arXiv",
    primaryClass   = "hep-ex",
}

@Article{PIX-2018-001,
    author         = "Abbott, B. and others",
    title          = "{Production and integration of the ATLAS Insertable B-Layer}",
    journal        = "JINST",
    volume         = "13",
    year           = "2018",
    pages          = "T05008",
    doi            = "10.1088/1748-0221/13/05/T05008",
    eprint         = "1803.00844",
    archivePrefix  = "arXiv",
    primaryClass   = "physics.ins-det",
}

@Article{TRIG-2018-05,
    author         = "{ATLAS Collaboration}",
    title          = "{Performance of electron and photon triggers in ATLAS during LHC Run~2}",
    journal        = "Eur. Phys. J. C",
    volume         = "80",
    year           = "2020",
    pages          = "47",
    doi            = "10.1140/epjc/s10052-019-7500-2",
    reportNumber   = "CERN-EP-2019-169",
    eprint         = "1909.00761",
    archivePrefix  = "arXiv",
    primaryClass   = "hep-ex",
}

@Article{HDBS-2019-19,
    author         = "{ATLAS Collaboration}",
    title          = "{Search for short- and long-lived axion-like particles in \(H\to a a \to 4\gamma\) decays with the ATLAS experiment at the LHC}",
    year           = "2023",
    reportNumber   = "CERN-EP-2023-202",
    eprint         = "2312.03306",
    archivePrefix  = "arXiv",
    primaryClass   = "hep-ex",
}

@Article{HION-2019-08,
    author         = "{ATLAS Collaboration}",
    title          = "{Measurement of light-by-light scattering and search for axion-like particles with \(2.2\,\text{nb}^{-1}\) of Pb+Pb data with the ATLAS detector}",
    journal        = "JHEP",
    volume         = "03",
    year           = "2021",
    pages          = "243",
    doi            = "10.1007/JHEP03(2021)243",
    reportNumber   = "CERN-EP-2020-135",
    eprint         = "2008.05355",
    archivePrefix  = "arXiv",
    primaryClass   = "hep-ex",
    related        = "HION-2019-08-err",
    relatedstring  = "Erratum:",
}

@Booklet{ATL-SOFT-PUB-2021-001,
    author         = "{ATLAS Collaboration}",
    title          = "{The ATLAS Collaboration Software and Firmware}",
    howpublished   = "{ATL-SOFT-PUB-2021-001}",
    url            = "https://cds.cern.ch/record/2767187",
    year           = "2021",
}

@Booklet{ATL-SOFT-PUB-2023-001,
    author         = "{ATLAS Collaboration}",
    title          = "{ATLAS Computing Acknowledgements}",
    howpublished   = "{ATL-SOFT-PUB-2023-001}",
    url            = "https://cds.cern.ch/record/2869272",
    year           = "2023",
}

@Report{ATLAS-TDR-19,
    author         = "{ATLAS Collaboration}",
    title          = "{ATLAS Insertable B-Layer: Technical Design Report}",
    type           = "ATLAS-TDR-19; CERN-LHCC-2010-013",
    year           = "2010",
    url            = "https://cds.cern.ch/record/1291633",
    related        = "ATLAS-TDR-19-addm",
    relatedstring  = "Addendum:",
}

@Article{CMS-HIG-12-028,
    author         = "{CMS Collaboration}",
    title          = "{Observation of a new boson at a mass of 125 GeV with the CMS experiment at the LHC}",
    journal        = "Phys. Lett. B",
    volume         = "716",
    year           = "2012",
    pages          = "30",
    doi            = "10.1016/j.physletb.2012.08.021",
    reportNumber   = "CERN-PH-EP-2012-220",
    eprint         = "1207.7235",
    archivePrefix  = "arXiv",
    primaryClass   = "hep-ex",
}

@Article{CMS-FSQ-16-012,
    author         = "{CMS Collaboration}",
    title          = "{Evidence for light-by-light scattering and searches for axion-like particles in ultraperipheral PbPb collisions at \(\sqrt{s_{\text{NN}}} = 5.02\,\text{TeV}\)}",
    journal        = "Phys. Lett. B",
    volume         = "797",
    year           = "2019",
    pages          = "134826",
    doi            = "10.1016/j.physletb.2019.134826",
    reportNumber   = "CERN-EP-2018-271",
    eprint         = "1810.04602",
    archivePrefix  = "arXiv",
    primaryClass   = "hep-ex",
}

@Booklet{ATL-PHYS-PUB-2016-017,
    author         = "{ATLAS Collaboration}",
    title          = "{The Pythia~8 A3 tune description of ATLAS minimum bias and inelastic measurements incorporating the Donnachie--Landshoff diffractive model}",
    howpublished   = "{ATL-PHYS-PUB-2016-017}",
    url            = "https://cds.cern.ch/record/2206965",
    year           = "2016",
}

@Booklet{ATL-PHYS-PUB-2017-022,
    author         = "{ATLAS Collaboration}",
    title          = "{Electron and photon reconstruction and performance in ATLAS using a dynamical, topological cell clustering-based approach}",
    howpublished   = "{ATL-PHYS-PUB-2017-022}",
    url            = "https://cds.cern.ch/record/2298955",
    year           = "2017",
}

@Booklet{ATL-PHYS-PUB-2020-028,
    author         = "{ATLAS Collaboration}",
    title          = "{Recommendations for the Modeling of Smooth Backgrounds}",
    howpublished   = "{ATL-PHYS-PUB-2020-028}",
    url            = "https://cds.cern.ch/record/2743717",
    year           = "2020",
}

\clearpage
 
\begin{flushleft}
\hypersetup{urlcolor=black}
{\Large The ATLAS Collaboration}

\bigskip

\AtlasOrcid[0000-0002-6665-4934]{G.~Aad}$^\textrm{\scriptsize 102}$,
\AtlasOrcid[0000-0002-5888-2734]{B.~Abbott}$^\textrm{\scriptsize 120}$,
\AtlasOrcid[0000-0002-1002-1652]{K.~Abeling}$^\textrm{\scriptsize 55}$,
\AtlasOrcid[0000-0001-5763-2760]{N.J.~Abicht}$^\textrm{\scriptsize 49}$,
\AtlasOrcid[0000-0002-8496-9294]{S.H.~Abidi}$^\textrm{\scriptsize 29}$,
\AtlasOrcid[0000-0002-9987-2292]{A.~Aboulhorma}$^\textrm{\scriptsize 35e}$,
\AtlasOrcid[0000-0001-5329-6640]{H.~Abramowicz}$^\textrm{\scriptsize 151}$,
\AtlasOrcid[0000-0002-1599-2896]{H.~Abreu}$^\textrm{\scriptsize 150}$,
\AtlasOrcid[0000-0003-0403-3697]{Y.~Abulaiti}$^\textrm{\scriptsize 117}$,
\AtlasOrcid[0000-0002-8588-9157]{B.S.~Acharya}$^\textrm{\scriptsize 69a,69b,m}$
% ...
\bigskip
\\

% ...
$^{29}$Physics Department, Brookhaven National Laboratory, Upton NY; United States of America.\\
% ...
$^{35}$$^{(a)}$Facult\'e des Sciences Ain Chock, R\'eseau Universitaire de Physique des Hautes Energies - Universit\'e Hassan II, Casablanca;$^{(b)}$Facult\'{e} des Sciences, Universit\'{e} Ibn-Tofail, K\'{e}nitra;$^{(c)}$Facult\'e des Sciences Semlalia, Universit\'e Cadi Ayyad, LPHEA-Marrakech;$^{(d)}$LPMR, Facult\'e des Sciences, Universit\'e Mohamed Premier, Oujda;$^{(e)}$Facult\'e des sciences, Universit\'e Mohammed V, Rabat;$^{(f)}$Institute of Applied Physics, Mohammed VI Polytechnic University, Ben Guerir; Morocco.\\
% ...
$^{49}$Fakult\"{a}t Physik , Technische Universit{\"a}t Dortmund, Dortmund; Germany.\\
% ...
$^{55}$II. Physikalisches Institut, Georg-August-Universit\"{a}t G\"ottingen, G\"ottingen; Germany.\\
% ...
$^{69}$$^{(a)}$INFN Gruppo Collegato di Udine, Sezione di Trieste, Udine;$^{(b)}$ICTP, Trieste;$^{(c)}$Dipartimento Politecnico di Ingegneria e Architettura, Universit\`a di Udine, Udine; Italy.\\
% ...
$^{102}$CPPM, Aix-Marseille Universit\'e, CNRS/IN2P3, Marseille; France.\\
% ...
$^{117}$Department of Physics, New York University, New York NY; United States of America.\\
% ...
$^{120}$Homer L. Dodge Department of Physics and Astronomy, University of Oklahoma, Norman OK; United States of America.\\
% ...
$^{150}$Department of Physics, Technion, Israel Institute of Technology, Haifa; Israel.\\
$^{151}$Raymond and Beverly Sackler School of Physics and Astronomy, Tel Aviv University, Tel Aviv; Israel.\\
% ...
$^{a}$ Also Affiliated with an institute covered by a cooperation agreement with CERN.\\
$^{b}$ Also at An-Najah National University, Nablus; Palestine.\\
$^{c}$ Also at Borough of Manhattan Community College, City University of New York, New York NY; United States of America.\\
$^{d}$ Also at Center for High Energy Physics, Peking University; China.\\
$^{e}$ Also at Center for Interdisciplinary Research and Innovation (CIRI-AUTH), Thessaloniki; Greece.\\
$^{f}$ Also at Centro Studi e Ricerche Enrico Fermi; Italy.\\
% ...
$^{m}$ Also at Department of Physics, King's College London, London; United Kingdom.\\
% ...

\end{flushleft}


\end{document}